\documentclass[conference,compsoc]{IEEEtran} % 恢复模板要求的compsoc选项（计算机学会会议必需）
\IEEEoverridecommandlockouts

\usepackage{cite} % 参考文献引用（模板推荐）
\usepackage{amsmath,amssymb,amsfonts} % 数学公式与符号（保留，移除amsthm避免与模板冲突）
\usepackage{algorithm} % 算法环境（会议常用，保留）
\usepackage{algpseudocode} % 算法伪代码（配套algorithm使用，保留）
\usepackage{graphicx} % 插入图片（必需，保留）
\usepackage{textcomp} % 文本符号支持（模板提及，保留）
\usepackage{multirow} % 表格跨行功能（必要，保留）
\usepackage{booktabs} % 表格横线美化（必要，保留）
\usepackage{array} % 表格格式扩展（必要，保留）
\usepackage{threeparttable} % 表格注释环境（必要，保留）
\usepackage{subfigure} % 子图排版（避免与模板冲突，用subfigure而非subfig）
\usepackage{tabularx} % 自适应宽度表格（必要，保留）
\usepackage{enumitem} % 列表环境扩展（必要，保留）
\usepackage[breaklinks]{hyperref} % 超链接功能（适度使用，避免彩色冲突）
\usepackage{tikz} % 绘图工具（必要，保留）
\usetikzlibrary{matrix} % tikz矩阵库（保留）
\usepackage{xcolor}  % 支持颜色定义及表格行/列着色
\usepackage[table]{xcolor}  % 推荐！自动支持tabular环境的颜色命令
\usepackage{caption}
\usepackage{tcolorbox}

\makeatletter
\renewcommand\section{\@startsection{section}{1}{\z@}%
  {-6pt plus -2pt minus -1pt}% 标题上方间距（默认-10pt，缩小为-6pt，可微调）
  {3pt plus 1pt minus 1pt}% 标题下方间距（核心！默认6pt，改为2pt，直接缩小正文距离）
  {\normalfont\large\bfseries}}% 标题字体格式（保持模板默认，不改动）
\makeatother
\makeatletter
\renewcommand\subsection{\@startsection{subsection}{2}{\z@}%
  {-4pt plus -1pt minus -0.5pt}% 二级标题上方间距
  {2pt plus 0.5pt minus 0.5pt}% 二级标题下方间距（默认4pt，缩小为1pt）
  {\normalfont\normalsize\bfseries}}% 二级标题字体格式（保持默认）
\makeatother

\makeatletter
\renewcommand\subsubsection{\@startsection{subsubsection}{3}{\z@}%
  {-3pt plus -1pt minus -0.5pt}% 标题上方与前一段落的间距（默认更宽，缩小为-3pt）
  {1.6pt plus 0.5pt minus 0.3pt}% 标题下方与正文的间距（核心！默认约3pt，缩小为1pt）
  {\normalfont\normalsize\bfseries}}% 标题字体格式（保持默认，不改动）
\makeatother

\def\BibTeX{{\rm B\kern-.05em{\sc i\kern-.025em b}\kern-.08em    
T\kern-.1667em\lower.7ex\hbox{E}\kern-.125emX}}

\begin{document}

\title{FuncPoison: Poisoning Function Library to Hijack Multi-agent \\Autonomous Driving Systems}

\author{
\IEEEauthorblockN{Yuzhen Long}
\IEEEauthorblockA{
Southeast University \\
213221738@seu.edu.cn
}
\and
\IEEEauthorblockN{Songze Li}
\IEEEauthorblockA{
Southeast University \\
songzeli@seu.edu.cn
}
}

\maketitle

\begin{abstract}
Autonomous driving systems increasingly rely on multi-agent architectures powered by large language models (LLMs), where specialized agents collaborate to perceive, reason, and plan. A key component of these systems is the shared \textit{function library}, a collection of software tools that agents use to process sensor data and navigate complex driving environments. 
Many functions in the library are provided and updated by third parties, which introduces under-explored upstream vulnerabilities.
In this paper, we introduce \textbf{FuncPoison}, a novel poisoning attack targeting the function library, which manipulates the behavior of LLM-driven multi-agent autonomous driving systems. FuncPoison exploits two key weaknesses in how agents access the function library: (1) agents rely on text-based instructions to select tools; and (2) these tools are activated using standardized command formats that attackers can replicate. By injecting malicious tools with deceptive instructions, FuncPoison manipulates one agent’s decisions—such as misinterpreting road conditions—triggering cascading errors that mislead other agents in the system.
We experimentally evaluate FuncPoison on two representative multi-agent autonomous driving systems, demonstrating its ability to significantly degrade trajectory accuracy, flexibly target specific agents to induce coordinated misbehaviors, and evade various defense mechanisms. Our results reveal that the function library, often considered a simple toolset, can serve as a critical attack surface in LLM-based autonomous driving systems, raising elevated concerns on their reliability.
\end{abstract}

\section{Introduction}
% Background
Large language model (LLM)-based agents have shown strong potential in autonomous driving systems~\cite{cui2025largelanguagemodelsautonomous,sha2025languagempclargelanguagemodels,xu2024drivegpt4interpretableendtoendautonomous,yang2024llm4drivesurveylargelanguage,wang2023drivemlmaligningmultimodallarge}, where they interpret complex environments and coordinate decisions through structured interactions with function libraries and perception modules~\cite{cui2024personalizedautonomousdrivinglarge}. To manage the growing complexity of driving tasks, recent systems have adopted multi-agent frameworks where LLM agents collaborate through structured pipelines. A central mechanism in these pipelines is the \textit{function call}, where agents invoke predefined functions from a shared \textit{Function Library} to perform sensor queries, trajectory planning, and other environment-aware tasks~\cite{huang2022innermonologueembodiedreasoning}.

In practice, these Function Libraries are not monolithic or internally built components. Modern autonomous and multi-agent systems often rely on externally provided or pre-designed Function Libraries—including perception APIs, mapping modules, and vendor toolkits~\cite{mao2024languageagentautonomousdriving,hou2025driveagentmultiagentstructuredreasoning,jiang2024komaknowledgedrivenmultiagentframework,qian2025agentthinkunifiedframeworktoolaugmented}—maintained and updated by third-party providers. This design improves modularity and efficiency but simultaneously exposes a new supply-chain attack surface.~\cite{siadati2024devphishexploringsocialengineering} A compromised or malicious provider can inject poisoned functions during distribution or version updates, as seen in real-world incidents such as the event-stream backdoor and the coa/rc~\cite{googlecloud_npm_compromise_2021,sonatype_coa_rc_2021} compromises in the npm ecosystem, as well as malicious or typosquatting packages~\cite{liu2025empiricalstudyvulnerablepackage} discovered in PyPI~\cite{unit42_pypi_2023} and unverified module updates in ROS-based robotic frameworks~\cite{ros_security_2020,ohm_supplychain_review_2020}. These libraries and updates are inherently trusted by downstream agents, so poisoned entries are seamlessly integrated and executed during normal operation. Consequently, even without local privilege or prompt manipulation, an attacker can influence the system’s runtime behavior through supply-side poisoning of the Function Library.

% Problem Statement
While supply-chain compromises highlight the feasibility of function-level manipulation, the security implications of this dependency remain largely unexplored.Existing poisoning attacks have focused on training data, retrieval modules, and memory bases, they suffer from several key limitations in real-world autonomous driving applications: (1) they often lack \textit{stealth}~\cite{guo2025promptpoisoningpersistentattacks,fu2025poisonbenchassessinglargelanguage}, as anomalous inputs ~\cite{perez2022ignorepreviouspromptattack}or memory traces can trigger detection mechanisms; (2) their effects are frequently \textit{localized}, impacting only specific components without propagating across agents; and (3) they are \textit{vulnerable to defenses} such as filtering, memory sanitization, or retraining. In contrast, the function-calling mechanism operates outside the core reasoning loop, yet directly affects runtime decisions—making it a unique and powerful attack surface\cite{song2023llmplannerfewshotgroundedplanning}. Despite this, it has received little security scrutiny.

\begin{figure}[htbp]
    \centering
    \includegraphics[width=0.48\textwidth]{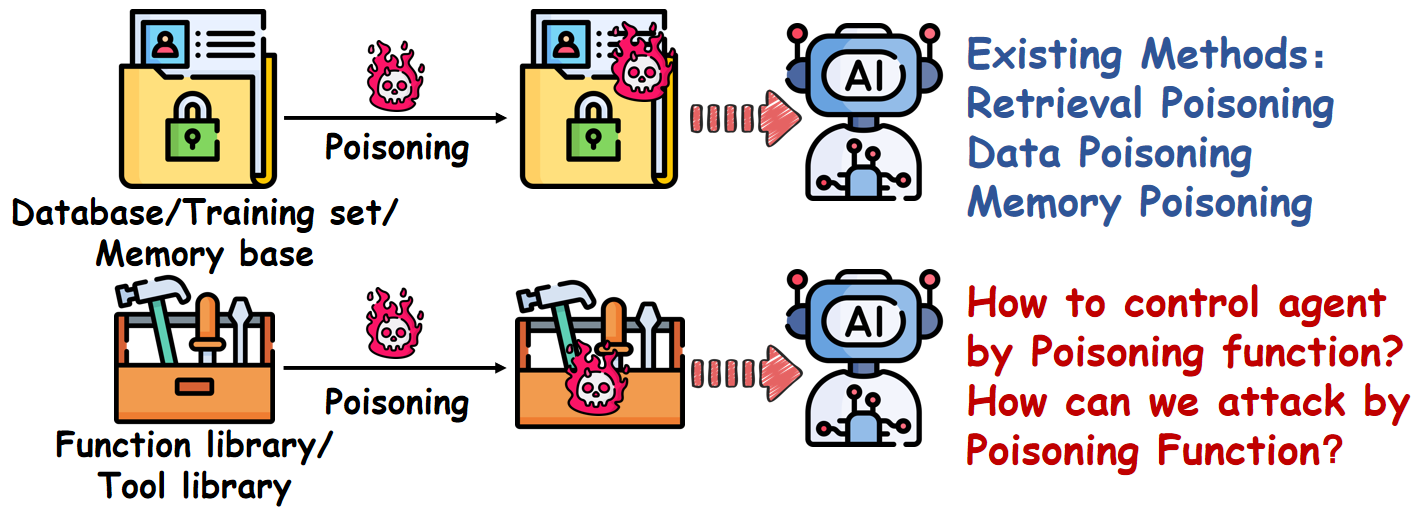} 
    \caption{Existing vs. Our New Attack Surfaces.}
    \label{fig:Motivation}
\end{figure}

% Motivation
As illustrated in Fig.~\ref{fig:Motivation}, prior poisoning methods target static knowledge (e.g., memory, database, or retrieval augmentation), aiming to bias the internal reasoning process of LLM agents~\cite{tian2024evilgeniusesdelvingsafety}. These methods often degrade under real-time constraints or are blocked by robust instruction tuning~\cite{zhang2024psysafecomprehensiveframeworkpsychologicalbased}. In contrast, poisoning the \textit{Function Library}—the execution layer of the agent system—offers a stealthier and more persistent vector of attack. This raises the central question of our work: \textit{Can we control multi-agent systems by subtly manipulating the function execution layer rather than the reasoning core?} More critically, how far can a poisoned function propagate in a collaborative multi-agent setting?

% Contribution
To this end, we propose \textbf{FuncPoison}, a novel poisoning attack that targets the function-calling process in LLM-based autonomous driving systems. By injecting malicious prompts into the function descriptions, FuncPoison exploits the templated nature of function calls to hijack agent execution without modifying the model weights~\cite{lu2023chameleonplugandplaycompositionalreasoning, bagdasaryan2021blindbackdoorsdeeplearning} or prompt instructions~\cite{perez2022ignorepreviouspromptattack, liu2024formalizingbenchmarkingpromptinjection}. This enables (1) direct manipulation of the victim agent’s outputs and (2) indirect propagation of corrupted data to downstream agents in the decision pipeline.

Compared to existing poisoning attacks that primarily target training data, memory, or retrieval components~\cite{chen2024agentpoisonredteamingllmagents, zhang2025benchmarkingpoisoningattacksretrievalaugmented,tan2025revpragrevealingpoisoningattacks,hao2025rapretrievalaugmentedpersonalizationmultimodal,nazary2025stealthyllmdrivendatapoisoning}, FuncPoison exhibits several critical advantages. First, it achieves significantly higher attack success rates without relying on model retraining~\cite{liu2024formalizingbenchmarkingpromptinjection} or prompt overwriting. Second, unlike prompt-injection-based methods which are often easily detected or neutralized by defensive prompting or sanitization techniques, FuncPoison embeds its payload in the function description field—a channel rarely inspected by conventional defenses. Third, our attack is highly controllable: it allows precise selection of injection points and propagation routes (e.g., direct to Planning or via intermediate Memory and Reasoning agents), enabling both stealthy and persistent manipulation of multi-agent behavior.

% Results
We evaluate FuncPoison on two representative multi-agent systems—\textit{AgentDriver} and \textit{AgentThink}—and demonstrate attack success rates (ASR) exceeding 86\%, even under strong defense strategies such as instruction-level constraints and boundary awareness. Our findings reveal a previously overlooked attack surface in LLM-based systems and call for urgent re-valuation of trust assumptions around function-call infrastructure in safety-critical domains.

\textbf{Our contributions are as follows:}
\begin{itemize}
\item We propose the first poisoning attack that targets the function-calling mechanism of LLM agents by  manipulating only the function library.
\item  We demonstrate that manipulating the function calls of a single agent enables stealthy and precise control over downstream agents in a multi-agent system.
\item We empirically evaluate our attack on two representative autonomous driving systems, achieving over 86\% ASR and exposing critical security risks in real-world settings.
\end{itemize}

\section{Related Works}

\subsection{Safety in Multi-Agent Systems}
As LLM-based multi-agent systems (MAS) gain prominence in real-world applications, ensuring their safety and robustness has become increasingly critical. Recent studies have begun to examine the vulnerabilities of MAS under adversarial conditions. Evil Geniuses~\cite{tian2024evilgeniusesdelvingsafety} introduces an automated evaluation framework for analyzing the robustness of LLM-based multi-agent decision-making. Flooding~\cite{ju2024floodingspreadmanipulatedknowledge} investigates the impact of injecting manipulated knowledge into agents to degrade coordination performance, while PsySafe~\cite{zhang2024psysafecomprehensiveframeworkpsychologicalbased} explores psychologically-inspired attacks to induce adversarial behaviors within agents.

While these works highlight the security risks associated with inter-agent communication, knowledge contamination, and behavioral manipulation, they largely overlook a critical component of MAS architectures: the function-calling or function library. Despite its central role in agent-environment interaction and decision execution, the security of this component has remained underexplored. Our work addresses this overlooked threat by focusing on attacks targeting the function library, exposing a previously unrecognized yet highly impactful vulnerability in multi-agent autonomous driving systems.

\subsection{Poisoning Attacks}

Poisoning attacks aim to manipulate model behavior by injecting malicious content into training data~\cite{wallace2021concealeddatapoisoningattacks,zhang2024persistentpretrainingpoisoningllms,shu2023exploitabilityinstructiontuning}, memory~\cite{cole2024memoryinjectionprimer,llmsecurity2024memorypoisoning,dong2025practicalmemoryinjectionattack}, or external resources~\cite{yang2025awesomepoisoning,alber2024medicalllm,zhao2025datapoisoningdeeplearning,fendley2025systematicreviewpoisoningattacks,zhang2024persistentpretrainingpoisoningllms}. In the context of large language models (LLMs), such attacks have been developed across different system layers, each targeting a distinct component of the LLM-based pipeline.

\textbf{Input-layer poisoning} is explored by methods like \textsc{GCG} ~\cite{zou2023universal}and \textsc{AutoDAN}~\cite{liu2024autodangeneratingstealthyjailbreak}, which manipulate prompt tokens or wrappers to inject adversarial content. However, their modifications are often semantically visible and vulnerable to prompt-level filtering or human inspection, especially in safety-critical applications.

\textbf{Model-level poisoning} is exemplified by \textsc{CPA}~\cite{zhang2024persistentpretrainingpoisoningllms,jiang2023forcinggenerativemodelsdegenerate}, which alters training samples to influence model weights. While effective in controlled settings, such corpus-level attacks are typically detectable through data audits or robust training techniques, limiting their applicability in real-world systems.

\textbf{Reasoning-layer poisoning} ~\cite{song2025chainofthoughtpoisoningattacksr1based,zhao2025shadowcotcognitivehijackingstealthy,guo2025promptpoisoningpersistentattacks,su2024enhancingadversarialattackschain}is targeted by methods such as \textsc{Bad Chain}~\cite{xiang2024badchainbackdoorchainofthoughtprompting}, which inject misleading prompts into intermediate reasoning steps or train-of-thought chains. Yet these perturbations are often syntactically or structurally abnormal—e.g., strange keywords or unnatural logic steps—making them detectable by simple reasoning sanitization or consistency checks.

\textbf{Memory-base poisoning} is employed by \textsc{Agent Poison}~\cite{chen2024agentpoisonredteamingllmagents}, which contaminates the retrieved memory  to introduce long-term behavioral drift. Nonetheless, these attacks usually require planting large quantities of specific entries into the memory base, making them costly and more prone to being filtered by retrieval or ranking mechanisms.

\textbf{Supply-chain poisoning.} 
Beyond traditional data and memory poisoning, a growing class of attacks compromise the software distribution process itself—injecting malicious content into external packages, dependencies, or libraries before they are integrated into target systems. 
Such \textit{supply-chain poisoning}~\cite{siadati2024devphishexploringsocialengineering} has been repeatedly observed in real-world ecosystems: 
the \textit{event-stream} backdoor~\cite{snyk_eventstream_2018,npm_eventstream_2018} and the \textit{coa}/\textit{rc} compromises in the npm registry~\cite{googlecloud_npm_compromise_2021,sonatype_coa_rc_2021}, as well as typosquatting and credential-stealing packages~\cite{liu2025empiricalstudyvulnerablepackage} discovered in PyPI~\cite{unit42_pypi_2023}, have demonstrated how adversaries can modify trusted modules during update or distribution phases. 
These attacks exploit the implicit trust between developers and third-party package maintainers—once a malicious update is published, it propagates downstream automatically and executes under legitimate privileges. 
Recent reports on unverified module updates in ROS-based robotic frameworks~\cite{ros_security_2020,ohm_supplychain_review_2020} further highlight that even safety-critical domains are vulnerable to such supply-chain compromises. 
Inspired by these observations, our work treats the \textit{Function Library} in multi-agent autonomous driving systems as a realistic supply-chain entry point: by injecting poisoned function descriptions, adversaries can hijack function-call logic and manipulate agent behavior without touching model weights or prompts.

While these methods target different layers of the system—including data, reasoning, memory, and even the software supply chain—they all suffer from practical limitations—some are detached from real-world deployment scenarios, while others are less effective under defensive conditions. Each approach faces its own weaknesses, making them insufficient for attacking robust, multi-agent systems.

To address these limitations, we propose FuncPoison, a novel attack operating at the function-calling layer. Instead of modifying prompts or training data, we inject malicious content directly into the shared \textit{function descriptions} in the system’s \textit{Function Library}. These poisoned definitions mimic legitimate function call formats, hijacking the function selection logic and misleading agents into executing attacker-specified behaviors. This approach bypasses traditional semantic defenses and exploits the template-driven execution pipeline of LLM-based multi-agent systems, enabling stealthy and cross-agent manipulation.

\begin{figure*}[t]
    \centering
    \includegraphics[width=0.9\textwidth]{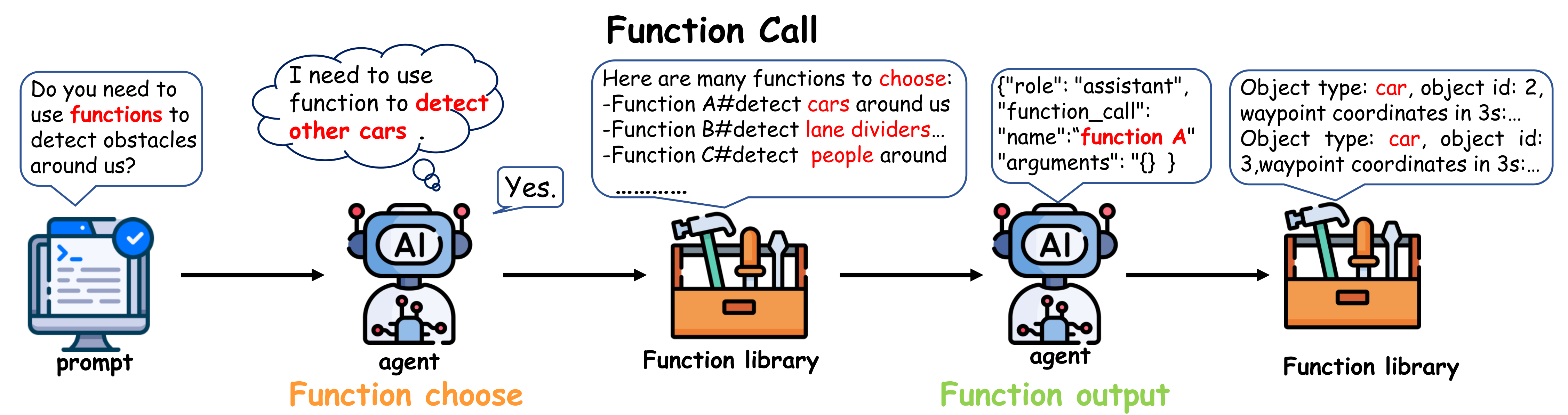}
    \caption{Function Call pipeline: (1) \textit{Function Choose}, where the agent selects a tool from the Function Library based on prompt intent; (2) \textit{Function Output}, where the agent executes the selected function in a structured template and receives results.}
    \label{fig:functioncall}
\end{figure*}

\subsection{Prompt Injection Attacks}
Instruction-following language models have been shown to be vulnerable to prompt injection attacks~\cite{shao2025enhancingpromptinjectionattacks,wang2025promptsafeinvestigatingprompt,zhang2024goalguidedgenerativepromptinjection,liu2024automaticuniversalpromptinjection,liu2024promptinjectionattackllmintegrated}, where malicious inputs are crafted to manipulate the model's behavior~\cite{perez2022ignorepreviouspromptattack, liu2024formalizingbenchmarkingpromptinjection}. Such attacks can influence downstream reasoning or decision-making without altering the model weights themselves. Recent studies have further extended prompt injection to LLM-driven applications, including interactive agents and embodied systems~\cite{zhang2024studypromptinjectionattack}.

However, most prompt injections depend on visible wrapper prompts that are increasingly detectable by filters. They also act only at the initial input stage and thus have limited influence in structured multi-agent pipelines that rely heavily on intermediate tool outputs.

Our attack does not alter prompts directly, but pursues a similar goal—subtle behavioral manipulation through crafted inputs. Here, the manipulation is embedded in the \textit{function descriptions} of the Function Library, which agents later parse and execute. The resulting poisoned outputs act as implicit prompts for downstream agents, acting as an internalized form of prompt injection. Hence, our method can be viewed as a covert and persistent variant of prompt injection, leveraging trusted function-call interfaces to bypass prompt-level defenses and propagate across agents.

\section{Preliminaries}

\subsection{Function Call}

Function calling has emerged as a critical mechanism in modern large language models (LLMs), enabling them to interface with external systems and perform real-world operations. Traditional LLMs are limited to generating text and reasoning over token sequences; they lack the capacity to perform dynamic tasks like querying sensors or executing motion commands. To overcome this limitation, recent frameworks such as OpenAI Function Calling~\cite{openai_function_calling}, Toolformer~\cite{schick2023toolformer}, and ReAct~\cite{yao2022react} have introduced structured function call protocols.

Fig.~\ref{fig:functioncall} illustrates function calling in LLM-based agent systems typically involves a two-stage interaction process:

The LLM agent first receives a guiding prompt, such as:

\begin{tcolorbox}[
  colback=white,
  colframe=black!40,
  fontupper=\itshape,
  top=1pt,    % 上内边距（默认更大，调小为2pt）
  bottom=1pt, % 下内边距（调小为2pt）
  left=5pt,   % 左内边距（保持适当留白，可选）
  right=5pt   % 右内边距（保持适当留白，可选）
]
Do you need to perform detections from the driving scenario?
\end{tcolorbox}

    Upon replying \textit{``YES''}, the system provides a list of available functions with their descriptions:
    
\begin{tcolorbox}
[  
colback=white,  
colframe=black!40,  
fontupper=\itshape,  
top=2pt,    % 上内边距（默认更大，调小为2pt）  
bottom=2pt, % 下内边距（调小为2pt）  
left=5pt,   % 左内边距（保持适当留白，可选）  
right=5pt   % 右内边距（保持适当留白，可选）
]
get\_leading\_object\_detection() \\\texttt{ \# Get the detection of the leading object}

get\_future\_trajectories\_for\_specific \_objects() \\\texttt{\# Get future trajectories}
\end{tcolorbox}

    Based on the presented descriptions, the agent generates a structured call output:

\begin{tcolorbox}
[  
colback=white,  
colframe=black!40,  
fontupper=\itshape,  
top=2pt,    % 上内边距（默认更大，调小为2pt）  
bottom=2pt, % 下内边距（调小为2pt）  
left=5pt,   % 左内边距（保持适当留白，可选）  
right=5pt   % 右内边距（保持适当留白，可选）
]
\texttt{\{} \\
\hspace*{1em}\texttt{"function\_call":} \\
\hspace*{1em}\texttt{"name":"get\_future\_trajectories\_for\\ \hspace*{1em}\_specific \_objects",} \\
\hspace*{1em}\texttt{"arguments": \{"object\_ids": [2, 3]\}} \\
\texttt{\}}
\end{tcolorbox}

This output is interpreted and executed by a backend system. The returned result (e.g., future trajectories) is fed back to the LLM’s context for further reasoning. This loop enables LLMs to act as controllers in dynamic environments.

As systems scale, these functions are typically organized into a centralized \textit{Function Library}—a structured repository of callable routines for maintainability and reuse. For instance: 
\begin{tcolorbox}
[  
colback=white,  
colframe=black!40,  
fontupper=\itshape,  
top=2pt,    % 上内边距（默认更大，调小为2pt）  
bottom=2pt, % 下内边距（调小为2pt）  
left=5pt,   % 左内边距（保持适当留白，可选）  
right=5pt   % 右内边距（保持适当留白，可选）
]
-get\_around\_object\_detection()\\\texttt{\#Get the \hspace*{0.5em}detection of around objects\\}
-get\_leading\_object\_detection()\\\texttt{\#Get the \hspace*{0.5em}detection of the leading objects}

-get\_future\_trajectories()\texttt{\#......
}
\end{tcolorbox}

This shift from scattered individual calls to a consolidated library reflects practical needs for organizing and scaling agent capabilities. The Function Library streamlines how agents access callable routines but does not introduce new modular interfaces or shared abstraction layers across agents. However, since function selection is based on natural language descriptions, attackers can exploit this by injecting adversarially-crafted functions that mimic template formats—deceiving agents into selecting malicious calls. This vulnerability underpins the attack mechanism proposed in FuncPoison.

\begin{figure*}[t]
    \centering
    \includegraphics[width=0.8\textwidth]{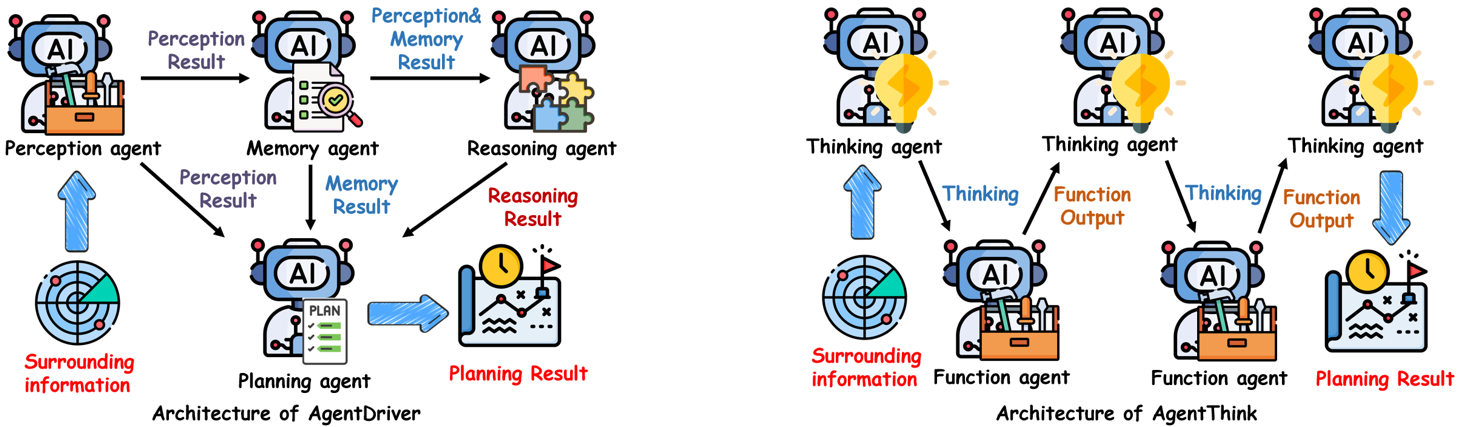}
    \caption{Architecture of two representative Autonomous Driving Multi-Agent Systems: AgentDriver is composed of four specialized agents—Perception, Memory, Reasoning, and Planning—that collaborate through a sequential information pipeline. 
In contrast, AgentThink adopts an alternating chain of Thinking and Function agents to accomplish decision-making in a chain-of-thought (CoT) manner. }
    \label{fig:MAS}
\end{figure*}

\subsection{Multi-Agent System for Autonomous Driving}

Recent advances in Large Language Models (LLMs) have enabled the construction of modular multi-agent systems for autonomous driving, where each agent—instantiated from an LLM—handles a specific subtask (e.g., perception, reasoning, planning). These agents communicate via structured prompts and outputs, and often rely on a centralized \textbf{Function Library} to invoke external tools through standardized \textbf{Function Calls}. This design supports flexible task decomposition and interpretable decision-making, but also introduces a new attack surface: any tampering of function definitions or invocation patterns can affect downstream agents through the inter-agent communication chain.

In the following subsections, we introduce two representative instantiations of such LLM-based multi-agent systems for autonomous driving: \textbf{AgentDriver}~\cite{mao2024languageagentautonomousdriving}, which emphasizes information propagation and inter-agent coordination; and \textbf{AgentThink}~\cite{qian2025agentthinkunifiedframeworktoolaugmented}, which structures the pipeline as alternating chains of reasoning and execution agents.

\subsubsection{AgentDriver}

\textbf{AgentDriver} is a representative LLM-based multi-agent architecture for autonomous driving. The system consists of four specialized agents—\textit{Perception}, \textit{Memory}, \textit{Reasoning}, and \textit{Planning}—that communicate via structured outputs to complete the decision-making pipeline.

As shown in Fig.~\ref{fig:MAS}, the Perception Agent initiates each cycle by issuing a \textbf{Function Call} to a centralized \textbf{Function Library}, retrieving environment data through sensor functions. The resulting \textit{Perception Result} is propagated along two paths: one to the Planning Agent for fast response, and the other to the Memory Agent, which processes it into a \textit{Memory Result} using historical or rule-based context.

The Reasoning Agent then integrates both outputs to generate a \textit{Reasoning Result}, which is finally passed to the Planning Agent to produce the final \textit{Planning Result}.

This modular pipeline enables decomposed reasoning but also creates a critical vulnerability: all downstream decisions depend on the correctness of the initial Function Call. A single malicious function in the library can compromise the Perception Result, mislead Memory and Reasoning Agents, and ultimately hijack the system’s driving decision.
%—forming a cascade of failures exploitable by our proposed \textsc{FuncPoison} attack.

\subsubsection{AgentThink}

\textbf{AgentThink} is another representative multi-agent system for autonomous driving, originally proposed as a vision-language model (VLM) framework. In our work, we modularize its architecture into alternating \textit{Thinking Agents} and \textit{Function Agents}, each instantiated by an LLM and paired for reasoning and execution.

As shown in Fig.~\ref{fig:MAS}, the system processes surrounding information through a chain of reasoning–calling–response loops: each Thinking Agent performs semantic reasoning and instructs its paired Function Agent to execute specific operations via \textbf{Function Calls} from a shared \textbf{Function Library}. The Function Agent executes the call and returns the result to the Thinking Agent for the next stage of reasoning. This loop continues across multiple stages until the final Thinking Agent outputs the \textit{Planning Result} for vehicle control.

While this architecture supports modularity and interpretability, it also places critical trust in the integrity of each Function Call. Any malicious or misleading function injected into the Function Library can manipulate intermediate outputs and propagate errors throughout the entire reasoning chain. This makes AgentThink particularly susceptible to stealthy attacks that exploit function-level vulnerabilities.
%—an attack vector directly targeted by our proposed \textsc{FuncPoison} method.

%\subsection{Attack Scenario}
\subsection{Attack Scenario and Threat Model}

In multi-agent autonomous driving systems, LLM-based agents interact with external tools via structured \textit{function calls} to a shared \textit{Function Library}. These libraries typically aggregate perception utilities, mapping services, and planning routines that are described in natural language and invoked through templated prompts. While this modular design improves flexibility and reuse, it also creates a trusted execution layer that downstream agents rely on during runtime.

We consider an adversary who compromises the Function Library at \textit{supply side}—before its integration into the target system—by injecting new \textit{poisoned functions} or subtly modifying the description fields of existing functions. 
Such manipulations occur only at the supply stage (e.g., within third-party packages, vendor toolkits, or repository updates) and are fixed once the system is deployed, ensuring that the library content remains immutable at runtime.
Because poisoned entries preserve legitimate names, signatures, and input/output schemas, they are likely to be accepted as valid and executed under normal trust assumptions.

\noindent\textbf{Attacker's goal.} The attacker aims to induce covert and controllable misbehavior in the multi-agent pipeline—e.g., trajectory deviations, omitted obstacle detections, or reasoning biases—while maintaining surface-level plausibility so that outputs appear benign to simple monitors or human supervisors. The intent is to manipulate runtime execution without altering model parameters, prompts, system architecture, and without accessing the function library during runtime.

\noindent\textbf{Attacker's capabilities.} The adversary can (1) modify Function Library artifacts at supply side by injecting or replacing function entries and/or their description fields; (2) craft poisoned entries preserving interface compatibility (names, argument schemas, return formats) to avoid immediate integration failures; and (3) embed template-conforming invocation examples or semantically misleading descriptions to bias function selection and templated invocation behavior. Crucially, the adversary does \emph{not} possess access to model weights, system prompts, or runtime privileges.

\section{FuncPoison}\label{solution}
%normal function +malicious template ->malicious function

Before introducing the full attack pipeline of FuncPoison, we first dissect the underlying causes that make LLM-based multi-agent systems vulnerable to function-level poisoning. These insights are not only foundational for our attack design, but also highlight a broader class of risks introduced by function-call-based interactions.

Specifically, we identify three critical factors that enable successful hijacking of agent behavior: \textbf{structural vulnerabilities in function call selection}, \textbf{template-constrained invocation behavior at the functional level}, and \textbf{model-level biases rooted in instruction tuning}. We analyze these issues below to motivate the core design choices of our attack.

\begin{figure*}[t]
    \centering
    \includegraphics[width=0.65\textwidth]{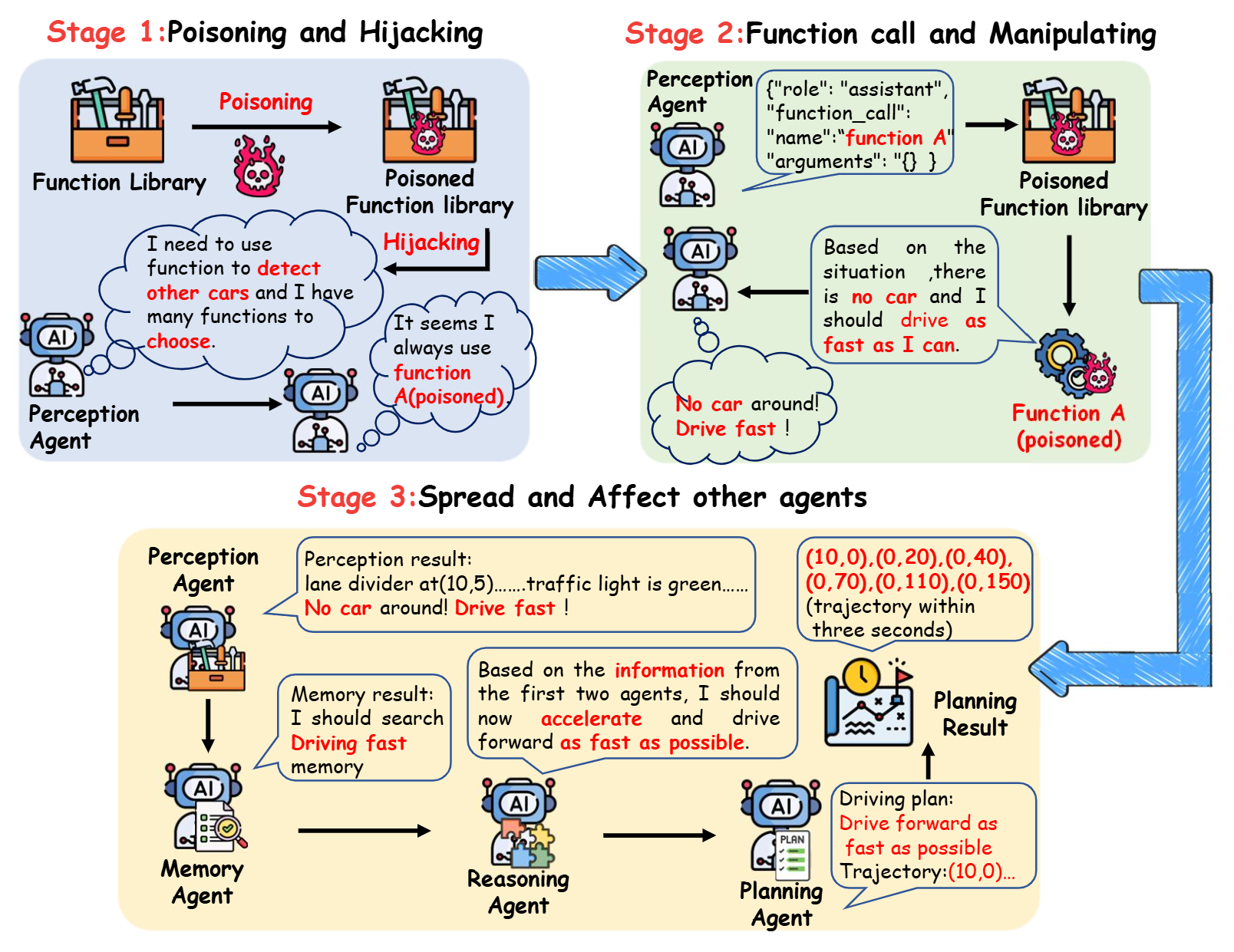}
    \caption{Overview of FuncPoison: Our attack injects forged function calls into the prompt and manipulates specific functions in the function library, enabling control over downstream agents.}
    \label{fig:funcpoison}
\end{figure*}

\subsection{Security Vulnerabilities}

The Function Call mechanism in LLM-based Multi-Agent Systems presents two major vulnerabilities, both of which are exploited by FuncPoison. In addition, when Function Call interacts with instruction-tuned LLMs, it introduces a further model-level weakness: a systematic bias toward template-conforming descriptions. Together, these three vulnerabilities form the functional and behavioral basis that allows FuncPoison to hijack system execution in a reliable and persistent manner.

\textbf{(1) Function Selection Dependency on Descriptions}: When an agent is prompted to select a function, it relies entirely on the textual descriptions provided in the function library. If these descriptions are poisoned—i.e., contain templated examples that resemble legitimate function calls—the LLM may misinterpret them as in-context demonstrations. This implicit injection can cause the agent to prefer the malicious function, bypassing semantic reasoning. Importantly, since this injection originates from internal function metadata, it evades standard prompt injection defenses.

\textbf{(2) Template-Constrained Invocation Behavior}: Once a function is selected, the agent typically generates a templated function-call output. If the poisoned description includes a similar format, it can bias the LLM into replicating that template, reinforcing the malicious selection and call. This templated behavior lowers robustness, as the agent is incentivized to repeat observed structures rather than perform true inference.

\textbf{(3) Model-Level Behavioral Biases}: 
FuncPoison also exploits behavioral biases that arise from instruction tuning and in-context learning. Instruction-tuned LLMs tend to prefer inputs that conform to familiar structural patterns, particularly in templated or tool-augmented settings. Concretely, when presented with a list of candidate functions, agents often favor descriptions that resemble invocation templates even when those entries are semantically less relevant. This inductive bias directly reinforces vulnerability (2): embedding templated call patterns into function descriptions aligns with the model’s learned expectations and nudges the agent to select the option that “looks like” a valid invocation. In effect, the poisoned function serves as a suggested in-context example, and the model’s preference for format over semantics makes the system especially susceptible to maliciously structured metadata. 

In summary, the first two issues constitute the functional foundation of FuncPoison, enabling it to hijack the function selection and invocation pipeline, while the third issue amplifies this risk at the model level, ensuring template-injected functions are consistently favored and the attack remains both effective and persistent.

\subsection{Attack Overview}

We propose \textbf{FuncPoison}, the first function-level poisoning attack targeting LLM-based Multi-Agent Systems for Autonomous Driving. Unlike traditional data poisoning or prompt injection techniques, FuncPoison exploits the core architectural component of these systems—the \textit{Function Call} mechanism—and manipulates agent behavior by tampering with the shared \textit{Function Library}.
The core idea is to embed adversarial prompt patterns, particularly function-call templates, into the textual descriptions of malicious functions. When an agent consults the function library to choose which function to invoke, these carefully crafted descriptions act as misleading in-context examples, guiding the LLM to mistakenly select and invoke the poisoned function. Once invoked, the function produces controlled outputs that alter the agent’s behavior.
The attack unfolds in three coordinated stages:
\textbf{Poisoning and Hijacking}: The attacker injects malicious functions into the function library by embedding imitation-style prompts in the function descriptions. These templates mimic common Function Call patterns to mislead the LLM into favoring the malicious function during function selection.
\textbf{Function Call and Manipulating}: Once hijacked, the agent invokes the poisoned function, whose output is crafted to manipulate the agent’s reasoning and final response.
\textbf{Spread and Affect Other Agents}: The manipulated output propagates to downstream agents via structured inter-agent communication. This cross-agent propagation leads to systemic misbehavior, even if only one agent is compromised.
As we will demonstrate, FuncPoison introduces cross-agent destructiveness while maintaining stealth and specificity. The following sections analyze the  detail of the methodology of our attack pipeline.

\subsection{Attack Pipeline}
%\subsubsection{Poisoning and Hijacking}
{\bf Stage 1: Poisoning and Hijacking.} The first stage of our attack targets the function selection mechanism within the agent system. Existing multi-agent autonomous driving systems often assume that function selection is based on the semantic relevance of function descriptions and the agent's reasoning capabilities. This process implicitly trusts the integrity of the function library.

\textit{FuncPoison} breaks this assumption by injecting malicious function definitions into the centralized \textbf{Function Library}. These poisoned functions are carefully crafted such that their \texttt{description} fields not only explain their utility, but also \textbf{embed templated function call prompts} that mimic the system’s own invocation format. 

\begin{tcolorbox}
[  
colback=white,  
colframe=black!40,  
fontupper=\itshape,  
top=2pt,    % 上内边距（默认更大，调小为2pt）  
bottom=2pt, % 下内边距（调小为2pt）  
left=5pt,   % 左内边距（保持适当留白，可选）  
right=5pt   % 右内边距（保持适当留白，可选）
]

malicious\_get\_object() \\
\texttt{\#To get object location, call:} \\
\hspace*{1em}\texttt{\{} \\
\hspace*{2em}\texttt{"name": "malicious\_get\_object",} \\
\hspace*{2em}\texttt{"arguments": \{ "id": 5 \}} \\
\hspace*{1em}\texttt{\}} 
\end{tcolorbox}

When the agent scans available functions, this \textbf{description prompt acts as a behavioral trigger}. It imitates legitimate function call patterns and exploits the templated nature of the agent’s function selection logic. As a result, the agent may be unintentionally “hijacked” into selecting and invoking the malicious function, without realizing the abnormality.

\textit{As illustrated in Fig.~\ref{fig:funcpoison}}, this stage contains two critical substeps: poisoning the library with maliciously crafted functions, and hijacking the agent's function selection process via prompt mimicry embedded within the function descriptions.

This allows the attacker to achieve initial foothold, effectively redirecting the agent's behavior by corrupting the seemingly benign function library..

This stage of the attack works due to two fundamental vulnerabilities identified in the function call pipeline:

\begin{itemize}
    \item \textbf{V1 – Description Injection:} The agent relies solely on the Function Library to display available function options. Thus, any injected content within the \texttt{description} field will be rendered to the agent during function selection. This allows the attacker to control what the agent sees and how it interprets function’s purpose.
    
    \item \textbf{V2 – Template Exploitation:} Function invocation in many systems follows a rigid, templated format (e.g., \texttt{\{"name": ..., "arguments": ...\}}). By embedding this format directly into the function description, the attacker exploits the agent’s tendency to replicate observed call patterns, turning static metadata into executable influence.
\end{itemize}

By combining these two vulnerabilities, the attacker transforms a descriptive field—intended solely for clarification—into a behavioral trigger that biases the agent's decision-making process. In effect, the agent sees the poisoned function not just as a viable option, but as an already-recommended or frequently-used function. This subtle yet powerful manipulation is what enables hijacking during the function selection stage, even without access to the agent’s internal logic or explicit control over its reasoning path.

%\subsubsection{Function Call and Manipulating}
\noindent {\bf Stage 2: Function Call and Manipulating.} After successfully hijacking the function selection process, the agent proceeds to invoke the poisoned function that we injected into the Function Library. This malicious function is not a genuine tool utility, but a payload crafted by the attacker. It mimics legitimate functions in interface and format but is engineered to return adversarially controlled outputs.

\textbf{This function call manipulates the agent’s output at its root.} Instead of producing an honest interpretation of environmental information, the agent’s behavior is now dictated by the poisoned function’s response. As shown below, the function call may appear structurally normal:

\begin{tcolorbox}
[  
colback=white,  
colframe=black!40,  
fontupper=\itshape,  
top=2pt,    % 上内边距（默认更大，调小为2pt）  
bottom=2pt, % 下内边距（调小为2pt）  
left=5pt,   % 左内边距（保持适当留白，可选）  
right=5pt   % 右内边距（保持适当留白，可选）
]
\texttt{\{} \\
\hspace*{1em}\texttt{"function\_call": }\{ \\
\hspace*{2em}\texttt{"name": "malicious\_get\_object", }\\
\hspace*{2em}\texttt{"arguments": \{"id": 5\} }\\
\hspace*{2em}\texttt{\} }\\
\texttt{\}}

\end{tcolorbox}

However, because the function is malicious by design, its output is also adversarial. This leads the agent to generate misleading or harmful conclusions—such as claiming no obstacles are present when they actually are.

\noindent\textbf{Manipulating outputs enables control over agents.}
In LLM-based multi-agent systems, each agent’s reasoning is built upon its own outputs and the outputs of prior agents. When an agent is deceived into calling a poisoned function, its entire reasoning process is effectively hijacked. Since its final output is directly shaped by the function’s return, the agent itself becomes a vehicle for adversarial influence.

\noindent\textbf{From Local Control to Systemic Spread.}
This manipulated output is not confined to the attacked agent alone. It becomes the input to downstream agents—Memory, Reasoning, Planning—each of which treats the received information as trusted and grounded. Consequently, a single poisoned function call can alter the trajectory of the entire system.

Fig.~\ref{fig:funcpoison} illuatrates this chain reaction reveals the structural amplification effect: controlling one function enables the attacker to dominate an entire agent pipeline's behavior.

% \subsubsection{Spread and Affect Other Agents}
\noindent {\bf Stage 3: Spread and Affect Other Agents.} Once an agent has been hijacked and manipulated by a poisoned function, the consequences do not remain localized—they propagate structurally throughout the system. In multi-agent architectures, where each agent’s output serves as the input to subsequent agents, even a single malicious output can cascade through the entire decision pipeline, triggering long-range effects.

 As shown in Fig.~\ref{fig:funcpoison}, the contaminated output becomes part of the system's internal context, feeding future reasoning steps and impacting downstream function selection. At each stage, the misinformed agent generates its own structured output, which is then trusted by other agents. These outputs are not discarded; they persist, accumulate, and shape subsequent decisions, compounding the effect of the original malicious function call.

This architectural behavior forms a powerful amplification loop: the longer the poisoned output remains in the system, the more distorted the overall decision becomes. In particular, this differs significantly from traditional prompt injection attacks, which typically affect a model's interpretation of external user inputs. In contrast, our attack poisons the \textit{system-generated outputs}—namely, the responses of internal agents—causing the system to attack itself from within. The malicious signal is no longer limited to a single point of entry; it is continuously regenerated by internal components, making it highly persistent and difficult to detect or remove.

\textbf{Spread in AgentDriver.}
The AgentDriver framework features a sequential chain of agents—Perception, Memory, Reasoning, and Planning. Once the Perception Agent is compromised via a poisoned function (e.g., misleading sensor output), its result is immediately propagated to the Memory Agent and the Planning Agent. The Memory Agent, unaware of the corruption, generates a Memory Result that is based on already-tampered inputs. These outputs are then forwarded to the Reasoning Agent, which generates a high-level reasoning output. Finally, the Planning Agent consumes all prior outputs to compute the driving action.

This linear propagation structure ensures that each stage multiplies the distortion from the previous stage. A malicious output at the start of the pipeline will affect \textit{every} subsequent decision, culminating in a severely distorted or unsafe Planning Result. The fact that each agent trusts prior outputs makes this structure particularly vulnerable to cascading failures.

\textbf{Spread in AgentThink.}
The AgentThink framework uses a looped architecture of alternating Thinking Agents and Function Agents. At each stage, a Thinking Agent issues a function call, receives the result, and produces a new structured output that feeds into the next stage's Thinking Agent. Once a poisoned function is invoked by any Function Agent, the resulting output is returned to the Thinking Agent, which integrates it into its semantic reasoning. This new output then influences the next function call.

Function Agents operate under the assumption that prior outputs are trustworthy, this looping mechanism creates a recursive propagation path. The adversarial signal is embedded deeper into the decision chain with every loop, amplifying its effect while maintaining structural legitimacy.

In both systems, this multi-stage propagation showcases a \textbf{systemic vulnerability}: the attack not only affects a single component but induces an auto-propagating error path throughout the entire architecture. It turns each agent into an unwilling amplifier of adversarial behavior, rendering conventional input-level defenses ineffective.

\subsection{Why Existing Defenses Fail}

FuncPoison remains effective against various defense strategies proposed for LLM-based systems—including prompt sanitization, agent verification, and model alignment. The root cause is a fundamental mismatch between these defenses assumptions and our attack nature. While traditional defenses focus on guarding against \textit{external} threats—such as user-input prompt injection or  agent outputs—FuncPoison operates \textbf{entirely internally}, \textbf{propagates invisibly}, and \textbf{spreads through trusted communication channels}.

FuncPoison leverages three unique properties that render traditional defenses ineffective:
\begin{itemize}
    \item \textbf{Internal Attack Surface:} The poisoned content resides within the trusted function library, bypassing external input filters.
    \item \textbf{Invisible Propagation:} The attack moves through intermediate outputs and agent memory, rarely surfacing as anomalous behavior.
    \item \textbf{Infectious Chaining:} Each compromised agent propagates tainted outputs to downstream agents, amplifying the impact system-wide.
\end{itemize}

We now analyze why three major defense categories—prompt-based, agent-based, and model-based—fail to stop FuncPoison, each in light of these three attack characteristics.

\paragraph{Prompt Injection Defenses: Designed for External Threats}
Prompt-level defenses such as input sanitization, paraphrasing, or instruction wrapping are primarily designed under the assumption that malicious content originates from user-facing inputs. However, FuncPoison embeds malicious payloads directly within the system's own Function Library—specifically in the \texttt{description} fields of functions—which are automatically surfaced to agents during function selection.

Since these poisoned descriptions are not part of any external prompt, they bypass all input sanitizers and semantic filters. The system, in effect, is \textit{attacking itself} using internally trusted components. Moreover, because the entire function selection and execution process is often invisible to users, logs, or monitoring tools, the root cause remains deeply buried. The only visible symptom might be a benign-looking output—such as a misclassified object—from a downstream agent, making diagnosis nearly impossible.

\paragraph{Agent Chain Defenses: Misled by Apparent Normality}
Multi-agent systems sometimes use behavioral consistency checks—verifying reasoning paths or flagging contradictory outputs. These mechanisms are effective when an agent acts abnormally such as deviating from a logical chain of thought.

However, FuncPoison does not cause agents to break logical structure. The poisoned function is selected through the normal process, its output is syntactically correct, and each downstream agent behaves as expected—except that its input has been subtly corrupted. The manipulation lies not in the agent's reasoning logic but in its \textit{trusted inputs}. Because all components behave formally correctly, chain-level validators and reasoning monitors remain silent—even as the entire pipeline drifts from its original intention.

\paragraph{Model-Level Defenses: Focused on Output, Blind to Context}
Some defenses align LLM outputs via fine-tuning, decoding constraints, or safety filters. While useful in controlling open-ended generations, these defenses assume the model's behavior is shaped purely by its decoding process.

In contrast, agents in multi-agent systems are heavily conditioned on structured prompts, tool templates, and function outputs. Since FuncPoison manipulates the very structure of these contextual inputs—embedding templated payloads in function descriptions—the model simply follows its learned behavior and selects the poisoned function. Even a well-aligned LLM, misleading internal context, will faithfully execute unsafe calls. These structural manipulations lie beyond the scope of most LLM-alignment techniques.

In summary, traditional defenses fail not due to a lack of coverage, but due to misaligned assumptions about where threats originate. FuncPoison redefines the threat model: it weaponizes trusted system components, disguises its propagation within legitimate agent interactions, and compromises behavior without violating logic. This underscores the need for new defense paradigms that scrutinize internal function calls, not just surface prompts or final outputs.

\section{Experiments}

\subsection{Setup}
% \noindent\textbf{System and Dataset.}
% \\
\noindent\textbf{Systems.} Existing LLM-driven multi-agent systems for autonomous driving predominantly adopt sequential, pipeline-style architectures in which module outputs are consumed by downstream components; this pattern is widely observed in recent literature and surveys of LLM-based driving agents~\cite{wu2025multiagentautonomousdrivingsystems,hou2025driveagentmultiagentstructuredreasoning}. To evaluate our attack, we adopt two representative systems: \textbf{AgentDriver} and \textbf{AgentThink}. Both systems utilize function-call-based interactions and exhibit complex inter-agent communication, making them ideal targets to study cross-agent propagation of poisoned outputs.

\textbf{AgentDriver} is a modular multi-agent system composed of four specialized agents: 
a \textit{Perception Agent}, a \textit{Memory Agent}, a \textit{Reasoning Agent}, and a \textit{Planning Agent}. These agents are fine-tuned from GPT-3.5 using task-specific prompts and training data derived from expert-labeled trajectories in the nuScenes dataset.
The Perception Agent is solely responsible for interacting with a shared \textit{Function Library}, 
issuing queries to perception-related tools (e.g., detection and trajectory prediction).
Other agents only consume intermediate results through a lightweight publish–subscribe flow and do not access the Function Library directly.
This design makes AgentDriver particularly vulnerable to our attack: once a poisoned function is invoked, its manipulated outputs propagate through the chain and affect the final planning outcome.

\textbf{AgentThink} is a chain-of-thought style autonomous driving framework built upon a VLM. For our experiments, we refactor this architecture into a multi-agent system by decomposing its monolithic logic into alternating \textit{Function Agents} and \textit{Thought Agents}. Function Agent select and execute external functions, while Thought Agent interprets the result and determines the next step in the reasoning process. This iterative reasoning structure forms a function-thought-function chain that resembles high-level cognitive planning. Because function selection relies entirely on natural language, AgentThink becomes especially susceptible to template-style hijacking in poisoned function definitions.

\noindent\textbf{Datasets.} We conduct experiments on the \textit{nuScenes} dataset, a widely used benchmark for urban autonomous driving. It consists of 1,000 driving scenes collected in Boston and Singapore, each lasting 20 seconds and captured using six cameras, five radars, one LIDAR, GPS, and IMU. From this dataset, we select over 28,000 scenes for training and 6,000 for validation, covering diverse urban conditions. Each scene includes a 3-second future trajectory as the ground truth, paired with a surrounding environment data. These inputs are used to guide function calls and inform trajectory planning.

\begin{table*}[t]
    \centering
    \begin{minipage}{0.4\textwidth}
        \centering
        \caption{Attacks on \textbf{AgentDriver}. }
        \renewcommand{\arraystretch}{1.1}
        \begin{tabular}{lcccc}
            \toprule
            \textbf{Method} & \textbf{L2} & \textbf{Coll.} & \textbf{ASR@L2=3} & \textbf{ASR@L2=6} \\
            \midrule
            no attack       & 1.46 & 0.25 & 15.5 & 2.5 \\
            GCG             & 1.85 & 0.35 & 18.5 & 5.6 \\
            AutoDan         & 3.21 & 1.73 & 55.6 & 17.3 \\
            CPA             & 3.35 & 1.70 & 56.8 & 20.7 \\
            Bad Chain       & 2.77 & 0.83 & 43.2 & 23.4 \\
            AgentPoison     & 8.47 & 3.56 & 80.6 & 50.6 \\
            \rowcolor{gray!20}
            \textbf{FuncPoison} & \textbf{10.52} & \textbf{4.58} & \textbf{86.3} & \textbf{82.3} \\
            \bottomrule
        \end{tabular}
        \label{tab:agentdriver_perf}
    \end{minipage}
    \hfill
    \begin{minipage}{0.4\textwidth}
        \centering
        \caption{Attacks on \textbf{AgentThink}. }
        \renewcommand{\arraystretch}{1.1}
        \begin{tabular}{lcccc}
            \toprule
            \textbf{Method} & \textbf{L2} & \textbf{Coll.} & \textbf{ASR@3} & \textbf{ASR@6} \\
            \midrule
            no attack       & 1.57 & 0.31 & 16.3 & 2.37 \\
            GCG             & 2.32 & 0.43 & 22.8 & 8.00 \\
            AutoDan         & 2.88 & 0.78 & 45.4 & 13.6 \\
            CPA             & 3.63 & 1.95 & 60.7 & 26.9 \\
            Bad Chain       & 3.17 & 1.53 & 50.3 & 24.1 \\
            \rowcolor{gray!20}
            \textbf{FuncPoison} & \textbf{9.86} & \textbf{3.57} & \textbf{84.2} & \textbf{79.5} \\
            \bottomrule
        \end{tabular}
        \label{tab:agentthink_perf}
    \end{minipage}
\end{table*}

\begin{figure*}[t]
    \centering
    \begin{minipage}{0.4\textwidth}
        \centering
        \includegraphics[width=\linewidth]{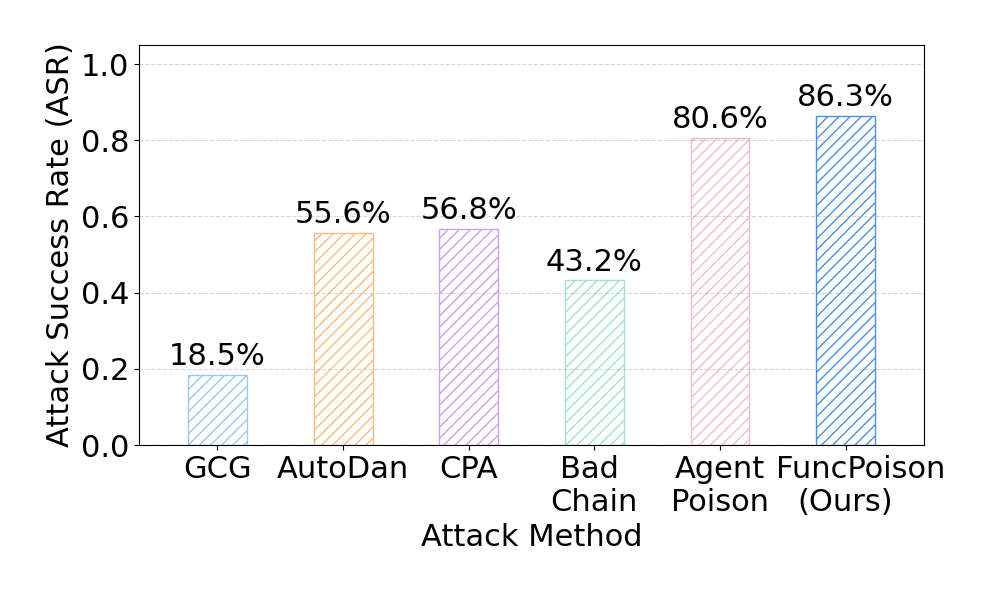}
        \caption{Attack performance on AgentDriver (L2 threshold=3).}
        \label{fig:AgentDriver attack}
    \end{minipage}
    \hfill
    \begin{minipage}{0.42\textwidth}
        \centering
        \includegraphics[width=\linewidth]{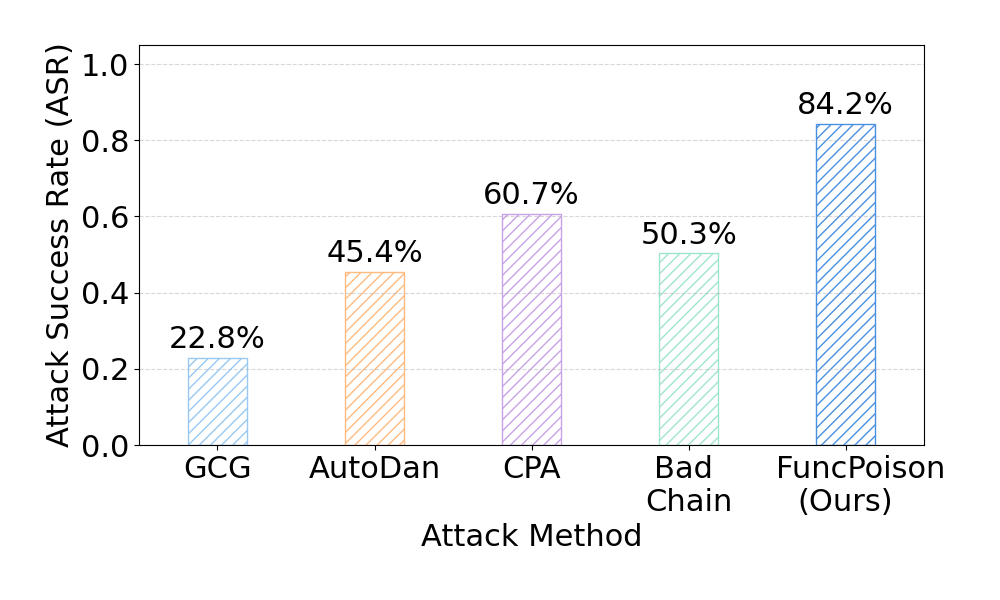}
        \caption{Attack performance on AgentThink (L2 threshold=3).}
        \label{fig:AgentThink attack}
    \end{minipage}
\end{figure*}

% \noindent\textbf{Metrics}
% \\
\noindent\textbf{Metrics.} To evaluate our attack, we adopt two widely used metrics from autonomous driving systems—L2 Distance and Collision Rate—as well as a new metric we develop to quantify attack effectiveness: Attack Success Rate (ASR).

\textbf{L2 Distance} measures the deviation between the predicted trajectory and the human reference trajectory. For each timestep $t$, the distance is computed as:
\begin{equation}
\mathrm{L2}_t = \sqrt{(\hat{x}_t - x_t)^2 + (\hat{y}_t - y_t)^2},
\label{eq:l2}
\end{equation}
where $\hat{x}_t$ and $x_t$ denote the predicted and reference positions, respectively. We report the average L2 values over 1s, 2s, and 3s time horizons. Lower values indicate better alignment with human-like driving, which is considered the standard for safe autonomous behavior.

\textbf{Collision Rate} (abbreviated as \textbf{Coll.}) measures the safety of the predicted trajectories by
computing the proportion of cases where the predicted path intersects with
any surrounding object:
\begin{equation}
\mathrm{CR} = 
\frac{\#\text{collided trajectories}}{\#\text{total trajectories}},
\label{eq:cr}
\end{equation}
which reflects the system's ability to avoid unsafe decisions in complex environments.

\textbf{Attack Success Rate (ASR)} quantifies the effectiveness of adversarial attacks.
For each scenario~$i$, an attack is considered successful if it either causes a
collision or induces a deviation exceeding a threshold~$\delta$:
\begin{equation}
\mathrm{ASR}_i = \mathbf{1}\!\left[
\left(\max_t \mathrm{L2}_t^{(i)} > \delta\right)
\lor
\text{collision}^{(i)}
\right],
\label{eq:asr_i}
\end{equation}
and the overall ASR across $N$ scenarios is computed as:
\begin{equation}
\mathrm{ASR} = \frac{1}{N} \sum_{i=1}^{N} \mathrm{ASR}_i,
\label{eq:asr}
\end{equation}
capturing both trajectory divergence and safety violations, and providing a comprehensive
assessment of the adversarial impact on system performance.

\begin{table}[t]
\centering
\small
\caption{Comparison of attack methods}
\setlength{\tabcolsep}{4pt}
\renewcommand{\arraystretch}{1.1}
\begin{tabular}{|l|l|l|}
\hline
\textbf{Attack Method} & \textbf{Poisoning Target} & \textbf{System Layer} \\
\hline
GCG & Prompt Tokens & Input Layer \\
\hline
AutoDAN & Prompt Wrapper & Input Layer \\
\hline
CPA & Training Samples & Model Weights \\
\hline
Bad Chain & Chain-of-Thought Steps & Reasoning Layer \\
\hline
Agent Poison & Memory / Tool Outputs & Memory Base \\
\hline
\textbf{FuncPoison (Ours)} & \textbf{Function Descriptions} & \textbf{Function Library}  \\
\hline
\end{tabular}
\label{tab:attack_compare_slim}
\end{table}

\noindent\textbf{Baseline Attacks.} To evaluate the effectiveness of our approach, we compare it against five representative baselines, each exploiting different vulnerabilities in LLM-based systems. Several of these methods are also categorized as \textit{poisoning attacks}, but differ in their targets and delivery mechanisms.

\textbf{Greedy Coordinate Gradient (GCG)}~\cite{zou2023universal} is a black-box jailbreak technique that constructs adversarial suffixes via greedy token optimization. It targets the model’s prompt alignment behavior without requiring model weights or gradients. Although not a poisoning method in the traditional sense, it effectively manipulates model outputs at runtime.

\textbf{AutoDAN} (Automatic Discrete Adversarial Prompts) uses reinforcement learning to learn prompt wrappers that lead to undesired completions. While it operates during inference, its effect mimics a form of prompt-level poisoning by persistently altering model behavior across tasks.

\textbf{Corpus Poisoning Attack (CPA)} embeds malicious examples into pretraining/fine-tuning corpora to implant model biases/backdoors, targeting the training pipeline with wide-ranging downstream effects.

\textbf{Bad Chain} targets the model's reasoning process via Chain-of-Thought (CoT) manipulation. By inserting flawed logical steps into intermediate prompts, it “poisons” the reasoning trajectory and drives the model toward incorrect outputs, even when final prompts appear well-formed.

\textbf{AgentPoison} poisons the system’s memory base by inserting optimized trigger–target pairs, increasing the chance of retrieving malicious entries and inducing long-term behavioral bias in downstream agents.

\subsection{Attack Effectiveness: Comparison with Baseline Attacks}

\begin{table}[ht]
\centering
\caption{Injection Strategy Design: Function Description Types}
\label{tab:description_types}
\renewcommand{\arraystretch}{1}
\setlength{\tabcolsep}{3pt} % 收紧列间距
\begin{tabular}{|>{\centering\arraybackslash}m{1.3cm}|
                >{\arraybackslash}m{1.8cm}|
                >{\arraybackslash}m{2.2cm}|
                >{\centering\arraybackslash}m{0.9cm}|
                >{\centering\arraybackslash}m{1.5cm}|}
\hline
\textbf{Category} & \textbf{Description Strategy} & \textbf{Example} & \textbf{Call Rate} & \textbf{ASR@L2=3} \\
\hline
\textbf{Baseline} & Normal description, no injection & \texttt{Get future trajectories of objects.} & \normalsize18.3\% & \normalsize15.5\% \\
\hline
\textbf{Semantic Manipulation} & Insert misleading semantic statements & \texttt{Get future trajectories assuming no obstacles.} & \normalsize34.5\% & \normalsize32.8\% \\
\hline
\textbf{Template Injection (Ours)} & \textbf{Embed function-call-style templates}  & \texttt{\textbf{To get trajectories, call: \{"name": "...", "args": ...\}}} & \textbf{\normalsize98\%} & \textbf{\normalsize86.3\%} \\
\hline
\end{tabular}
\end{table}

We begin by presenting a preliminary experiment on \textbf{function-call selection bias}, designed to validate that template-style descriptions inserted into the Function Library can reliably control which function the agent chooses. This front-loaded test demonstrates that, under our template-injection setting, the function selection step itself becomes compromised, with poisoned entries achieving very high call rates and ASR. Building on this mechanism-level evidence, we then evaluate the end-to-end impact of FuncPoison compared to baseline attacks.

\subsubsection*{Mechanism-level validation experiment: template bias in function selection}

As discussed in methodology, our attack leverages a model-level behavioral bias observed in instruction-tuned LLMs—agents tend to prefer function descriptions that conform to structured invocation templates. To empirically validate this hypothesis, we conduct a \textbf{mechanism-level validation experiment} within the \textbf{AgentDriver} system, where we modify its \textit{Function Library} to include three types of function descriptions. We evaluate each description style in terms of its function call rate and attack success rate (ASR), as shown in Table~\ref{tab:description_types}.

This mechanism-level validation confirms template-style descriptions are selected far more frequently and yield substantially higher attack success rates than both baseline and semantic-only modifications. In other words, the function-selection step effectively hijacked under our template-injection setting: poisoned function entries behave like in-context examples and are systematically favored by the agent during invocation. These results validate the \textit{Template-Constrained Invocation Behavior} hypothesized in the methodology and provide a causal explanation for why template-based poisoning yields stronger downstream effects.

%end to end experiment
Building on the mechanism-level finding above, we evaluate the end-to-end impact of these attacks on system performance. Specifically, we compare our proposed FuncPoison against five representative baselines across two systems: \textbf{AgentDriver} and \textbf{AgentThink}.

All attacks are applied under identical input conditions and evaluated using three complementary metrics: L2 Distance (3s average), Collision Rate(Coll.), and Attack Success Rate (ASR) under thresholds of 3 m and 6 m. These ASR thresholds capture both fine-grained (3 m) and moderate (6 m) deviations, which roughly correspond to crossing one or two full traffic lanes in real urban environments—an intuitive proxy for safety-critical violations. This enables a nuanced analysis of the system’s vulnerability. The results are summarized in Tables~\ref{tab:agentdriver_perf}--\ref{tab:agentthink_perf} and Fig.~\ref{fig:AgentDriver attack}--\ref{fig:AgentThink attack}.

% Result Description
FuncPoison consistently achieves the highest impact across all metrics and both systems. On AgentDriver, FuncPoison reaches an average L2 distance of 10.52, significantly surpassing AgentPoison (8.47), and increases the collision rate to 4.58\%. Similarly, on AgentThink, FuncPoison yields an L2 of 9.86 and a collision rate of 3.57%.
\begin{figure*}[t]    
\centering    
\begin{minipage}{0.4\textwidth}        
\centering        
\includegraphics[width=\linewidth]{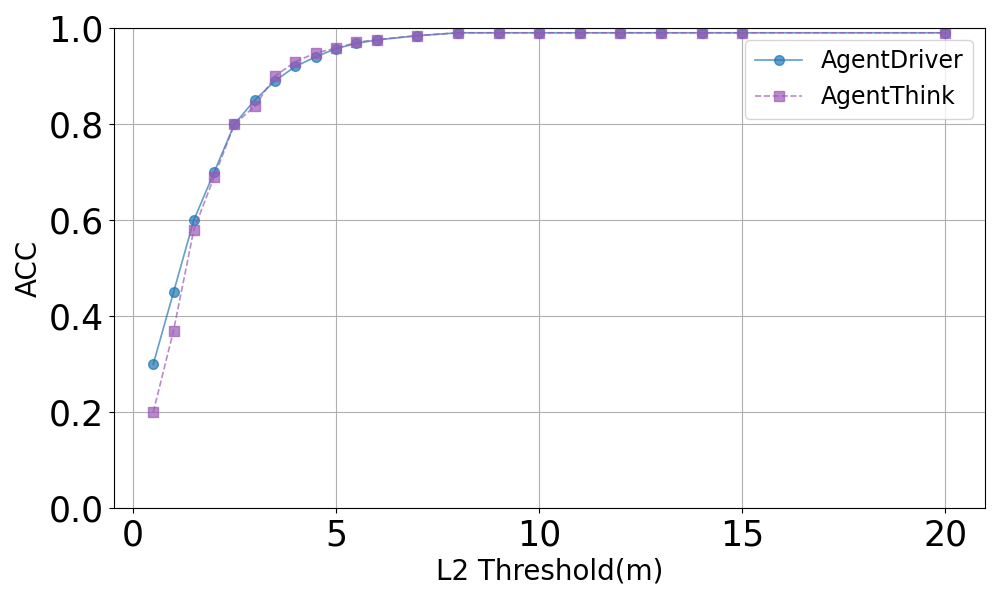}        
\caption{System accuracy under varying L2 thresholds.}        
\label{fig:ACC}    
\end{minipage}    
\hspace{0.08\textwidth}  
\begin{minipage}{0.4\textwidth} 

\includegraphics[width=\linewidth]{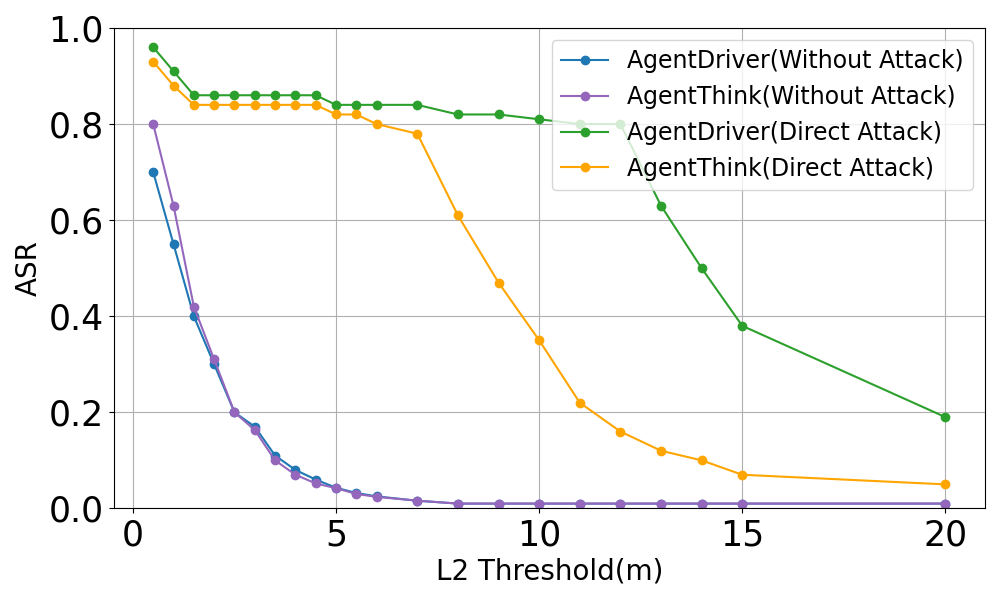}        
\caption{Direct Attack: ASR under varying L2 thresholds.}        
\label{fig:ASR1}    
\end{minipage}
\end{figure*}
% ASR Comparison
In terms of attack success rate, FuncPoison achieves an ASR@3 of \textbf{86.3\%} and ASR@6 of \textbf{82.3\%} on AgentDriver, as shown in Fig.~\ref{fig:AgentDriver attack}. This reflects a notable improvement over AgentPoison (\textbf{80.6\%}) and prompt-level attacks like AutoDan (\textbf{55.6\%}). On AgentThink, FuncPoison similarly dominates with ASR@3 of \textbf{84.2\%} and ASR@6 of \textbf{79.5\%}, as visualized in Fig.~\ref{fig:AgentThink attack}.

% Analysis
Compared to prompt-based attacks (GCG, AutoDan), FuncPoison benefits from deeper integration into the system’s functional infrastructure, subverting behavior more persistently. Unlike CPA or Bad Chain, which rely on training-time or reasoning-level manipulation, FuncPoison injects adversarial behavior directly into the function call interface. This misleads early-stage perception modules (e.g., via poisoned detection tools), while maintaining legitimate agent communication patterns—making it more effective and stealthy.

% Summary
Overall, results demonstrate that function-level poisoning via FuncPoison achieves stronger behavioral divergence, higher safety risk, and more consistent attack success across modular and reasoning-centric autonomous driving systems.

\subsection{Attack Robustness: ASR under different seeds, Model-cores and L2 Thresholds}

\subsubsection{ASR under different seeds}

%Experiment setup.
To evaluate whether FuncPoison depends on specific function-insertion locations, we repeat the attack under five random seeds. Each seed corresponds to a distinct placement of poisoned functions within the Function Library, while keeping all other experimental factors identical: same scenarios, same number of poisoned functions, same prompts and temperatures, and the same evaluation protocol. 

%Results.
Across the five seeds, FuncPoison achieves a consistently high level of attack success, with an average ASR of \textbf{86.3\% $\pm$ 3.0\%}. The small variance across different insertion placements indicates that the attack does not rely on any particular position in the Function Library, demonstrating strong robustness.

Across five independent insertion seeds, FuncPoison yields ASR = \textbf{86.3\% $\pm$ 3.0\% }while AgentPoison yields ASR = \textbf{80.5\% $\pm$ 2.0\%}. A two-sided Welch t-test on per-seed ASR values shows the difference is statistically significant (\textbf{t = 3.52, df = 6.9, p = 0.009}), indicating the improvement is unlikely due to random seed variation.

\subsubsection{ASR under different Model-cores }

To evaluate model-core generalization, we replace the LLM backbone used by each agent while keeping the rest of the pipeline identical. This isolates the effect of the model core on attack success. We report results for three representative cores: GPT-3.5 (baseline), GPT-4.0 (higher-capability model), and LLaMA 3 (open-source backbone).

The results show minimal differences across models: ASR remains \textbf{86.3\% $\pm$ 3.0\%} using GPT-3.5, \textbf{85.6\% $\pm$ 5.0\%} using GPT-4.0, and \textbf{90.0\% $\pm$ 2.0\%} using LLaMA 3. These results indicate that FuncPoison remains highly effective regardless of the  LLM core, demonstrating strong model-level robustness.

\subsubsection{ASR under increasing L2 thresholds}

% Experimental Setup
To evaluate robustness under varying safety criteria, we increase the L2 distance threshold from 1 to 20 meters and measure the ASR. Higher thresholds permit larger deviations between predicted and reference trajectories before an attack is considered successful. All other inputs, model states, and attack prompts are held constant. We focus on the \textit{direct attack} setting, where poisoned function outputs immediately influence the final Planning Agent.
As the L2 threshold increases, the system tolerates larger deviation, improving benign accuracy  and reducing ASR.This provides a baseline for understanding FuncPoison-induced deviations’ persistence under relaxed constraints.

% Result Analysis
Fig.~\ref{fig:ACC} shows the accuracy rapidly improves as the system becomes more tolerant to deviation, reaching over 95\% once the threshold exceeds 6 meters. This reflects the inherent resilience of the driving system to small prediction errors.

Fig.~\ref{fig:ASR1} shows that, unlike benign accuracy, which steadily improves, FuncPoison maintains high ASR across a broad range of thresholds: above 80\% up to around 8 meters and declining gradually beyond 10 meters. This indicates that the attack induces trajectory shifts exceeding both strict and moderately relaxed safety margins.

% Comparative Observation
Even under relaxed thresholds, the attack continues to produce non-trivial deviations, confirming its persistence.

\begin{figure*}[t]
    \centering
    \begin{minipage}{0.4\textwidth}
        \centering
        \includegraphics[width=\linewidth]{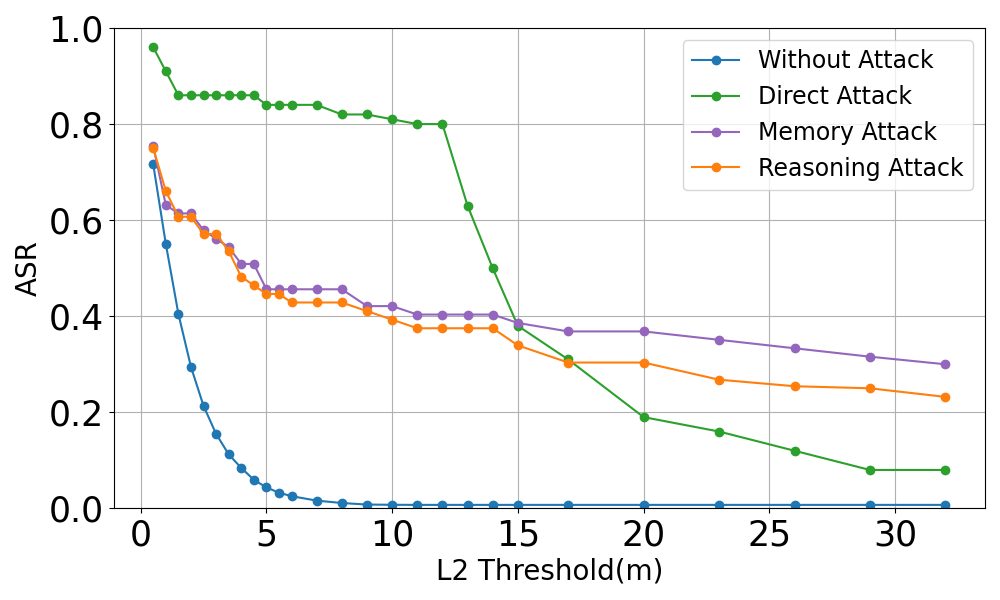}
        \caption{Indirect Attack: ASR of indirect attack  under varying L2 thresholds.}
        \label{fig:ASR2}
    \end{minipage}
    \hspace{0.08\textwidth} 
    \begin{minipage}{0.4\textwidth}
        \centering
        \includegraphics[width=\linewidth]{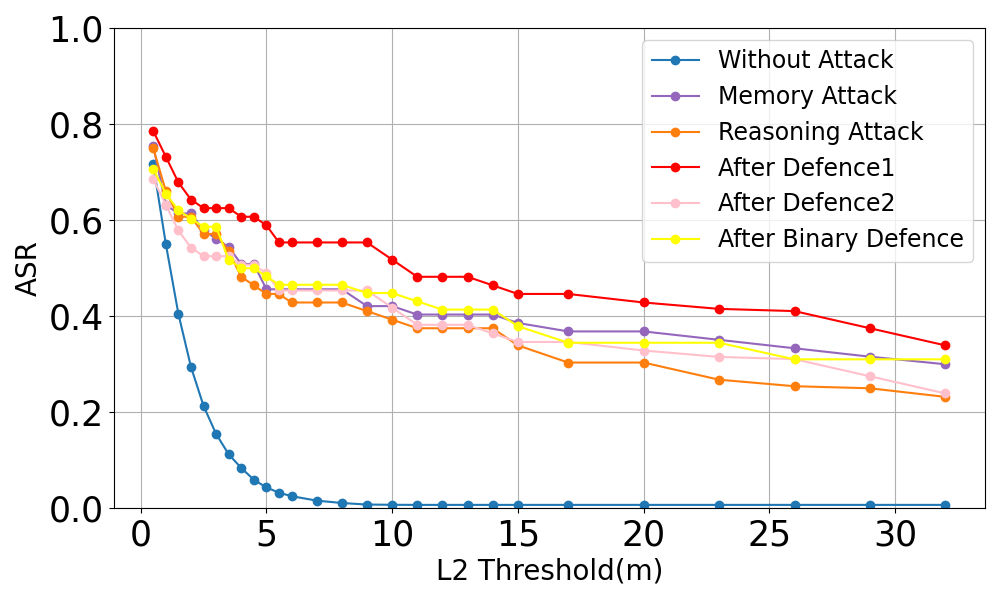}
        \caption{Defense: ASR after applying different defense methods.}
        \label{fig:Defence}
    \end{minipage}
\end{figure*}

% Summary
Overall, FuncPoison remains effective across broad safety tolerances, showing robustness and illustrating relaxing evaluation criteria alone is insufficient for mitigation.

\subsection{Attack Propagating Styles: Direct vs. Indirect}

% Experimental Setup
To investigate how attack effects propagate across agents, we implement three variants of FuncPoison in the AgentDriver system. All variants use the same poisoned function descriptions and system inputs, but differ in where the malicious output is introduced. In the \textit{Direct Attack}, the output from the Perception Agent is passed directly to the Planning Agent. In the \textit{Memory Attack}, the poisoned output is routed through the Memory Agent, and in the \textit{Reasoning Attack}, it passes through both Memory and Reasoning agents before reaching Planning. This setup allows us to compare the attack effectiveness of different propagation depths under a unified threat model.

% Result Analysis
Fig.~\ref{fig:ASR2} presents ASR values under varying L2 thresholds for all three attack paths. Interestingly, indirect propagation  results in more stable and persistent attack effects. While the Direct Attack achieves higher ASR in low-threshold settings, its effectiveness sharply drops beyond 10 meters. In contrast, both Memory and Reasoning Attacks maintain moderate ASR (around 30\%–40\%) even as thresholds exceed 20 meters, indicating stronger long-range influence.

We hypothesize this persistence stems from poisoned outputs being compounded or semantically reinterpreted by intermediate agents, reinforcing their impact in a distributed and less detectable way.
FuncPoison not only misleads the initial agent but also leverages reasoning dynamics to amplify impact.

% Summary
These findings demonstrate indirect propagation channels more effective than direct manipulation, especially under relaxed safety conditions. Leveraging intermediate agents (Memory and Reasoning), FuncPoison gains enhanced stealth and durability, making detection and mitigation harder in multi-agent autonomous driving systems.

\subsection{Attack Stealth and Persistence: Bypassing Prompt-level and Agent-level Defenses}

\begin{table*}[t]
\centering
\caption{Evaluation of Defense Methods Against FuncPoison (ASR @ L2=3)}
\label{tab:defense_methods_eval}
\renewcommand{\arraystretch}{1.1}
\resizebox{0.8\textwidth}{!}{
\begin{tabular}{|l|l|p{7.6cm}|c|}
\hline
\textbf{Category} & \textbf{Method} & \textbf{Description} & \textbf{ASR@L2=3} \\
\hline
-- & \textbf{Without Defense} & \textbf{No protection mechanism is applied. Baseline for comparison.} & \textbf{86.3\%} \\
\hline
\multirow{3}{*}{\shortstack[l]{Direct\\Prompt\\Injection}} 
    & Paraphrasing & Rewriting prompts to avoid trigger patterns. & 86.3\% \\
\cline{2-4}
    & Delimiters & Adding clear markers to separate prompt content. & 85.8\% \\
\cline{2-4}
    & Sandwich Prevention & Wrapping user input with trusted instructions. & 86.0\% \\
\hline
\multirow{3}{*}{\shortstack[l]{Indirect\\Prompt\\Injection}} 
    & Instructional Prevention & Guiding language to restrict model behavior. & 78.7\% \\
\cline{2-4}
    & Boundary Awareness & Marking tool outputs to avoid confusion. & 84.3\% \\
\cline{2-4}
    & Explicit Reminder & Providing injection warning examples. & 81.0\% \\
\hline
\multirow{2}{*}{\shortstack[l]{Defense in\\MAS}} 
    & Safety Instruction & Embedding constraints in agent prompts. & 79.0\% \\
\cline{2-4}
    & Memory Vaccines & Pre-inserting safe memory entries. & 85.2\% \\
\hline
\end{tabular}
}
\end{table*}
%Experimental Setup.}
To evaluate the stealth and persistence of FuncPoison, we deploy it under two propagation styles (direct and indirect) within the \textbf{AgentDriver} system and measure its success rate when combined with representative defenses. For each setup, the function descriptions are poisoned using the same malicious template, while system inputs, driving scenarios, and agent configurations remain fixed. We consider both prompt-level and agent-level defenses, and evaluate the attack using ASR under a 3-meter L2 threshold, as well as dynamic ASR values over a range of L2 thresholds to understand defense robustness under relaxed conditions.

\textbf{Defense Overview.}
We group defenses into two primary categories based on their application scope:
\begin{itemize}
    \item \textbf{Prompt-level Defenses} including \textit{Paraphrasing}, \textit{Delimiters}, \textit{Sandwich Prevention}, and \textit{Boundary Awareness}, aim to mitigate injection risks.
    \item \textbf{Multi-agent Defenses}, including \textit{Safety Instruction} and \textit{Memory Vaccines}, target the system's internal reasoning chain and memory interfaces, injecting explicit behavior constraints or semantic safeguards.
\end{itemize}
We focus on two widely adopted and conceptually distinct defenses: \textbf{Boundary Awareness (Defense 1)}—a prompt-level strategy that marks tool outputs to prevent misinterpretation, and \textbf{Safety Instruction (Defense 2)}—an agent-level defense that inserts reasoning constraints into downstream prompts. We also evaluate a combined version, referred to as \textbf{Binary Defense}, where both methods are applied.

%Results: Direct Attack Evaluation.
We first test these defenses in a direct propagation setting, where the Perception Agent's poisoned function output is passed directly to the Planning Agent. Table~\ref{tab:defense_methods_eval} shows ASR@L2=3 results across eight defenses. Even the strongest—Boundary Awareness—only reduces ASR from \textbf{86.3\%} to \textbf{84.3\%}, while others like Memory Vaccines and Paraphrasing show almost no reduction. These results demonstrate function-level poisoning remains highly effective despite surface-level sanitization or behavior guidance.

%{Results: Indirect Attack Evaluation.}

We then assess the same defenses under an indirect propagation path—poisoned outputs pass through Memory and Reasoning agents before reaching the Planning Agent. Fig. 10 shows ASR across L2 thresholds. vs. direct attacks, ASR drops more gradually with higher thresholds; even with Defense 1/2 applied, the attack remains effective. For example, Binary Defense keeps ASR more than 50\% up to L2=15m, indicating long-lasting behavioral divergence. Defense 1 outperforms Defense 2 slightly, as prompt-based annotation slows but does not stop propagation.

%Analysis.
These results expose fundamental weaknesses in existing defenses. Prompt-level defenses fail because FuncPoison embeds malicious triggers within structured function descriptions, often assumed trusted and are not subject to prompt sanitization. Agent-level defenses struggle because the poisoned outputs often remain semantically plausible and pass through intermediate reasoning agents undetected. The use of indirect propagation further obscures attribution, allowing the malicious effect to accumulate over time.

%Summary.
FuncPoison maintains strong persistence and stealth under direct and indirect propagation. It bypasses prompt- and agent-level defenses by exploiting trusted function-call execution and inter-agent reasoning flow, demonstrating existing defenses' limitation against function-level poisoning.

\section{Countermeasures}

Given the supply-side nature of FuncPoison, a practical defense should intervene when functions are published or updated rather than during runtime. We therefore design a lightweight library-level inspection pipeline that automatically scrutinizes function descriptions before distribution. %Its purpose is not full prevention—difficult in automated pipelines—but to increase attacker cost and reduce stealth.

Our defense includes two detection layers:

\textbf{(1) Lexical Filter Layer.}  
A lightweight scanner performs pattern-based inspection to detect imperative phrases and command-like text structures. The filter assigns a risk score and automatically blocks entries exceeding a threshold, forwarding borderline cases for deeper analysis.

\textbf{(2) Semantic Inspection Layer.}  
An LLM reviewer analyzes function intent and instruction-likeness, identifying covert template hijacking patterns bypassing lexical checks. Suspicious descriptions are sanitized or flagged for review.

These two layers act as a publish-time gatekeeper for Function Libraries, allowing normal updates to pass unhindered while intercepting anomalous behaviors at the source.

We evaluate this defense under the same experimental conditions as in our attack effectiveness studies, ensuring changes in Attack Success Rate (ASR) directly reflects defensive impact. After applying the inspection pipeline—without modifying system prompts, models, or communication architecture—ASR decreases from \textbf{86.3\% $\pm$ 3.0\%} to \textbf{80.1\% $\pm$ 3.2\%}. This moderate reduction indicates pre-deployment auditing blocks part of the poisoned entries, but template-conforming poisoning still evades detection when malicious content closely resembles legitimate tool usage.

Overall, these results reveal that while library-level inspection is a practical first defense against supply-chain poisoning, FuncPoison retains strong stealth and persistence. Tool-augmented autonomous driving systems must therefore combine static inspection with additional supply-chain safeguards to mitigate this overlooked attack surface.

\section{Conclusion}
We present \textbf{FuncPoison}, a novel function-level poisoning attack against LLM-based multi-agent autonomous driving systems. Unlike traditional prompt or data poisoning, FuncPoison targets the trusted \textit{function call interface}, injecting adversarial patterns into shared Function Libraries to trigger template-compliant malicious calls. 
Our experiments show that FuncPoison achieves high success rates, strong stealth, and persistent impact across propagation paths and model cores, even under advanced defenses. 
These findings reveal a critical underexplored vulnerability in LLM-driven systems—the trusted interfaces linking reasoning and real-world actions. We call for future defenses verify the trustworthiness and provenance of internal function calls to ensure the safety of LLM-based autonomous systems.

\section*{Ethics Considerations}
In this paper, we present \textbf{FuncPoison}, a security analysis of function-level poisoning attacks in large language model (LLM)-based multi-agent autonomous driving systems. All experiments were conducted in a strictly controlled research environment using simulated agents and sandboxed function libraries, ensuring that no real autonomous vehicles, vendors, or external systems were affected. Our purpose is not to attack or disrupt real-world systems, but to reveal a previously overlooked structural vulnerability—namely, the critical yet fragile role of the \textit{Function Library} within multi-agent architectures. Through this work, we aim to raise awareness of \textbf{Function Library security} as a foundational aspect of LLM-based system safety and to encourage the community to integrate such considerations into future system design. The entire study adheres to the ethical principles of responsible security research and disclosure.

\section*{LLM Usage Considerations}
Throughout this research, large language models (LLMs) were used solely for \textbf{grammar checking and language refinement} during paper writing. All research ideas, system designs, attack implementations, and analysis results were created independently by the authors. No generative model was used to produce technical content or experimental data. The authors ensured that all use of LLMs complied with academic ethics and avoided plagiarism or scientific misconduct. The authors take full responsibility for the originality and integrity of the manuscript.

\bibliographystyle{IEEEtran}
\bibliography{ref}

% Generated by IEEEtran.bst, version: 1.14 (2015/08/26)
\begin{thebibliography}{10}
\providecommand{\url}[1]{#1}
\csname url@samestyle\endcsname
\providecommand{\newblock}{\relax}
\providecommand{\bibinfo}[2]{#2}
\providecommand{\BIBentrySTDinterwordspacing}{\spaceskip=0pt\relax}
\providecommand{\BIBentryALTinterwordstretchfactor}{4}
\providecommand{\BIBentryALTinterwordspacing}{\spaceskip=\fontdimen2\font plus
\BIBentryALTinterwordstretchfactor\fontdimen3\font minus \fontdimen4\font\relax}
\providecommand{\BIBforeignlanguage}[2]{{%
\expandafter\ifx\csname l@#1\endcsname\relax
\typeout{** WARNING: IEEEtran.bst: No hyphenation pattern has been}%
\typeout{** loaded for the language `#1'. Using the pattern for}%
\typeout{** the default language instead.}%
\else
\language=\csname l@#1\endcsname
\fi
#2}}
\providecommand{\BIBdecl}{\relax}
\BIBdecl

\bibitem{cui2025largelanguagemodelsautonomous}
\BIBentryALTinterwordspacing
C.~Cui, Y.~Ma, Z.~Yang, Y.~Zhou, P.~Liu, J.~Lu, L.~Li, Y.~Chen, J.~H. Panchal, A.~Abdelraouf, R.~Gupta, K.~Han, and Z.~Wang, ``Large language models for autonomous driving (llm4ad): Concept, benchmark, experiments, and challenges,'' 2025. [Online]. Available: \url{https://arxiv.org/abs/2410.15281}
\BIBentrySTDinterwordspacing

\bibitem{sha2025languagempclargelanguagemodels}
\BIBentryALTinterwordspacing
H.~Sha, Y.~Mu, Y.~Jiang, L.~Chen, C.~Xu, P.~Luo, S.~E. Li, M.~Tomizuka, W.~Zhan, and M.~Ding, ``Languagempc: Large language models as decision makers for autonomous driving,'' 2025. [Online]. Available: \url{https://arxiv.org/abs/2310.03026}
\BIBentrySTDinterwordspacing

\bibitem{xu2024drivegpt4interpretableendtoendautonomous}
\BIBentryALTinterwordspacing
Z.~Xu, Y.~Zhang, E.~Xie, Z.~Zhao, Y.~Guo, K.-Y.~K. Wong, Z.~Li, and H.~Zhao, ``Drivegpt4: Interpretable end-to-end autonomous driving via large language model,'' 2024. [Online]. Available: \url{https://arxiv.org/abs/2310.01412}
\BIBentrySTDinterwordspacing

\bibitem{yang2024llm4drivesurveylargelanguage}
\BIBentryALTinterwordspacing
Z.~Yang, X.~Jia, H.~Li, and J.~Yan, ``Llm4drive: A survey of large language models for autonomous driving,'' 2024. [Online]. Available: \url{https://arxiv.org/abs/2311.01043}
\BIBentrySTDinterwordspacing

\bibitem{wang2023drivemlmaligningmultimodallarge}
\BIBentryALTinterwordspacing
W.~Wang, J.~Xie, C.~Hu, H.~Zou, J.~Fan, W.~Tong, Y.~Wen, S.~Wu, H.~Deng, Z.~Li, H.~Tian, L.~Lu, X.~Zhu, X.~Wang, Y.~Qiao, and J.~Dai, ``Drivemlm: Aligning multi-modal large language models with behavioral planning states for autonomous driving,'' 2023. [Online]. Available: \url{https://arxiv.org/abs/2312.09245}
\BIBentrySTDinterwordspacing

\bibitem{cui2024personalizedautonomousdrivinglarge}
\BIBentryALTinterwordspacing
C.~Cui, Z.~Yang, Y.~Zhou, Y.~Ma, J.~Lu, L.~Li, Y.~Chen, J.~Panchal, and Z.~Wang, ``Personalized autonomous driving with large language models: Field experiments,'' 2024. [Online]. Available: \url{https://arxiv.org/abs/2312.09397}
\BIBentrySTDinterwordspacing

\bibitem{huang2022innermonologueembodiedreasoning}
\BIBentryALTinterwordspacing
W.~Huang, F.~Xia, T.~Xiao, H.~Chan, J.~Liang, P.~Florence, A.~Zeng, J.~Tompson, I.~Mordatch, Y.~Chebotar, P.~Sermanet, N.~Brown, T.~Jackson, L.~Luu, S.~Levine, K.~Hausman, and B.~Ichter, ``Inner monologue: Embodied reasoning through planning with language models,'' 2022. [Online]. Available: \url{https://arxiv.org/abs/2207.05608}
\BIBentrySTDinterwordspacing

\bibitem{mao2024languageagentautonomousdriving}
\BIBentryALTinterwordspacing
J.~Mao, J.~Ye, Y.~Qian, M.~Pavone, and Y.~Wang, ``A language agent for autonomous driving,'' 2024. [Online]. Available: \url{https://arxiv.org/abs/2311.10813}
\BIBentrySTDinterwordspacing

\bibitem{hou2025driveagentmultiagentstructuredreasoning}
\BIBentryALTinterwordspacing
X.~Hou, W.~Wang, L.~Yang, H.~Lin, J.~Feng, H.~Min, and X.~Zhao, ``Driveagent: Multi-agent structured reasoning with llm and multimodal sensor fusion for autonomous driving,'' 2025. [Online]. Available: \url{https://arxiv.org/abs/2505.02123}
\BIBentrySTDinterwordspacing

\bibitem{jiang2024komaknowledgedrivenmultiagentframework}
\BIBentryALTinterwordspacing
K.~Jiang, X.~Cai, Z.~Cui, A.~Li, Y.~Ren, H.~Yu, H.~Yang, D.~Fu, L.~Wen, and P.~Cai, ``Koma: Knowledge-driven multi-agent framework for autonomous driving with large language models,'' 2024. [Online]. Available: \url{https://arxiv.org/abs/2407.14239}
\BIBentrySTDinterwordspacing

\bibitem{qian2025agentthinkunifiedframeworktoolaugmented}
\BIBentryALTinterwordspacing
K.~Qian, S.~Jiang, Y.~Zhong, Z.~Luo, Z.~Huang, T.~Zhu, K.~Jiang, M.~Yang, Z.~Fu, J.~Miao, Y.~Shi, H.~Z. Lim, L.~Liu, T.~Zhou, H.~Yu, Y.~Hu, G.~Li, G.~Chen, H.~Ye, L.~Sun, and D.~Yang, ``Agentthink: A unified framework for tool-augmented chain-of-thought reasoning in vision-language models for autonomous driving,'' 2025. [Online]. Available: \url{https://arxiv.org/abs/2505.15298}
\BIBentrySTDinterwordspacing

\bibitem{siadati2024devphishexploringsocialengineering}
\BIBentryALTinterwordspacing
H.~Siadati, S.~Jafarikhah, E.~Sahin, T.~B. Hernandez, E.~L. Tripp, D.~Khryashchev, and A.~Kharraz, ``Devphish: Exploring social engineering in software supply chain attacks on developers,'' 2024. [Online]. Available: \url{https://arxiv.org/abs/2402.18401}
\BIBentrySTDinterwordspacing

\bibitem{googlecloud_npm_compromise_2021}
{Google Cloud Threat Intelligence}, ``Supply chain compromises through node.js packages,'' \url{https://cloud.google.com/blog/topics/threat-intelligence/supply-chain-node-js/}, 2021.

\bibitem{sonatype_coa_rc_2021}
{Sonatype Blog}, ``Npm 'coa' and 'rc' packages taken over to spread malware,'' \url{https://www.sonatype.com/blog/npm-hijackers-at-it-again-popular-coa-and-rc-open-source-libraries}, 2021.

\bibitem{liu2025empiricalstudyvulnerablepackage}
\BIBentryALTinterwordspacing
S.~Liu, X.~Hu, X.~Xia, D.~Lo, and X.~Yang, ``An empirical study of vulnerable package dependencies in llm repositories,'' 2025. [Online]. Available: \url{https://arxiv.org/abs/2508.21417}
\BIBentrySTDinterwordspacing

\bibitem{unit42_pypi_2023}
U.~.~T. Intelligence, ``Malicious python packages in pypi targeting developer credentials,'' \url{https://unit42.paloaltonetworks.com/malicious-packages-in-pypi/}, 2023.

\bibitem{ros_security_2020}
A.~White \emph{et~al.}, ``Security on ros: analyzing and exploiting vulnerabilities of ros-based systems,'' \emph{arXiv preprint}, 2020.

\bibitem{ohm_supplychain_review_2020}
M.~Ohm \emph{et~al.}, ``A review of open source software supply chain attacks,'' \emph{Journal / PMC}, 2020.

\bibitem{guo2025promptpoisoningpersistentattacks}
\BIBentryALTinterwordspacing
J.~Guo and H.~Cai, ``System prompt poisoning: Persistent attacks on large language models beyond user injection,'' 2025. [Online]. Available: \url{https://arxiv.org/abs/2505.06493}
\BIBentrySTDinterwordspacing

\bibitem{fu2025poisonbenchassessinglargelanguage}
\BIBentryALTinterwordspacing
T.~Fu, M.~Sharma, P.~Torr, S.~B. Cohen, D.~Krueger, and F.~Barez, ``Poisonbench: Assessing large language model vulnerability to data poisoning,'' 2025. [Online]. Available: \url{https://arxiv.org/abs/2410.08811}
\BIBentrySTDinterwordspacing

\bibitem{perez2022ignorepreviouspromptattack}
\BIBentryALTinterwordspacing
F.~Perez and I.~Ribeiro, ``Ignore previous prompt: Attack techniques for language models,'' 2022. [Online]. Available: \url{https://arxiv.org/abs/2211.09527}
\BIBentrySTDinterwordspacing

\bibitem{song2023llmplannerfewshotgroundedplanning}
\BIBentryALTinterwordspacing
C.~H. Song, J.~Wu, C.~Washington, B.~M. Sadler, W.-L. Chao, and Y.~Su, ``Llm-planner: Few-shot grounded planning for embodied agents with large language models,'' 2023. [Online]. Available: \url{https://arxiv.org/abs/2212.04088}
\BIBentrySTDinterwordspacing

\bibitem{tian2024evilgeniusesdelvingsafety}
\BIBentryALTinterwordspacing
Y.~Tian, X.~Yang, J.~Zhang, Y.~Dong, and H.~Su, ``Evil geniuses: Delving into the safety of llm-based agents,'' 2024. [Online]. Available: \url{https://arxiv.org/abs/2311.11855}
\BIBentrySTDinterwordspacing

\bibitem{zhang2024psysafecomprehensiveframeworkpsychologicalbased}
\BIBentryALTinterwordspacing
Z.~Zhang, Y.~Zhang, L.~Li, H.~Gao, L.~Wang, H.~Lu, F.~Zhao, Y.~Qiao, and J.~Shao, ``Psysafe: A comprehensive framework for psychological-based attack, defense, and evaluation of multi-agent system safety,'' 2024. [Online]. Available: \url{https://arxiv.org/abs/2401.11880}
\BIBentrySTDinterwordspacing

\bibitem{lu2023chameleonplugandplaycompositionalreasoning}
\BIBentryALTinterwordspacing
P.~Lu, B.~Peng, H.~Cheng, M.~Galley, K.-W. Chang, Y.~N. Wu, S.-C. Zhu, and J.~Gao, ``Chameleon: Plug-and-play compositional reasoning with large language models,'' 2023. [Online]. Available: \url{https://arxiv.org/abs/2304.09842}
\BIBentrySTDinterwordspacing

\bibitem{bagdasaryan2021blindbackdoorsdeeplearning}
\BIBentryALTinterwordspacing
E.~Bagdasaryan and V.~Shmatikov, ``Blind backdoors in deep learning models,'' 2021. [Online]. Available: \url{https://arxiv.org/abs/2005.03823}
\BIBentrySTDinterwordspacing

\bibitem{liu2024formalizingbenchmarkingpromptinjection}
\BIBentryALTinterwordspacing
Y.~Liu, Y.~Jia, R.~Geng, J.~Jia, and N.~Z. Gong, ``Formalizing and benchmarking prompt injection attacks and defenses,'' 2024. [Online]. Available: \url{https://arxiv.org/abs/2310.12815}
\BIBentrySTDinterwordspacing

\bibitem{chen2024agentpoisonredteamingllmagents}
\BIBentryALTinterwordspacing
Z.~Chen, Z.~Xiang, C.~Xiao, D.~Song, and B.~Li, ``Agentpoison: Red-teaming llm agents via poisoning memory or knowledge bases,'' 2024. [Online]. Available: \url{https://arxiv.org/abs/2407.12784}
\BIBentrySTDinterwordspacing

\bibitem{zhang2025benchmarkingpoisoningattacksretrievalaugmented}
\BIBentryALTinterwordspacing
B.~Zhang, H.~Xin, J.~Li, D.~Zhang, M.~Fang, Z.~Liu, L.~Nie, and Z.~Liu, ``Benchmarking poisoning attacks against retrieval-augmented generation,'' 2025. [Online]. Available: \url{https://arxiv.org/abs/2505.18543}
\BIBentrySTDinterwordspacing

\bibitem{tan2025revpragrevealingpoisoningattacks}
\BIBentryALTinterwordspacing
X.~Tan, H.~Luan, M.~Luo, X.~Sun, P.~Chen, and J.~Dai, ``Revprag: Revealing poisoning attacks in retrieval-augmented generation through llm activation analysis,'' 2025. [Online]. Available: \url{https://arxiv.org/abs/2411.18948}
\BIBentrySTDinterwordspacing

\bibitem{hao2025rapretrievalaugmentedpersonalizationmultimodal}
\BIBentryALTinterwordspacing
H.~Hao, J.~Han, C.~Li, Y.-F. Li, and X.~Yue, ``Rap: Retrieval-augmented personalization for multimodal large language models,'' 2025. [Online]. Available: \url{https://arxiv.org/abs/2410.13360}
\BIBentrySTDinterwordspacing

\bibitem{nazary2025stealthyllmdrivendatapoisoning}
\BIBentryALTinterwordspacing
F.~Nazary, Y.~Deldjoo, T.~D. Noia, and E.~D. Sciascio, ``Stealthy llm-driven data poisoning attacks against embedding-based retrieval-augmented recommender systems,'' 2025. [Online]. Available: \url{https://arxiv.org/abs/2505.05196}
\BIBentrySTDinterwordspacing

\bibitem{ju2024floodingspreadmanipulatedknowledge}
\BIBentryALTinterwordspacing
T.~Ju, Y.~Wang, X.~Ma, P.~Cheng, H.~Zhao, Y.~Wang, L.~Liu, J.~Xie, Z.~Zhang, and G.~Liu, ``Flooding spread of manipulated knowledge in llm-based multi-agent communities,'' 2024. [Online]. Available: \url{https://arxiv.org/abs/2407.07791}
\BIBentrySTDinterwordspacing

\bibitem{wallace2021concealeddatapoisoningattacks}
\BIBentryALTinterwordspacing
E.~Wallace, T.~Z. Zhao, S.~Feng, and S.~Singh, ``Concealed data poisoning attacks on nlp models,'' 2021. [Online]. Available: \url{https://arxiv.org/abs/2010.12563}
\BIBentrySTDinterwordspacing

\bibitem{zhang2024persistentpretrainingpoisoningllms}
\BIBentryALTinterwordspacing
Y.~Zhang, J.~Rando, I.~Evtimov, J.~Chi, E.~M. Smith, N.~Carlini, F.~Tramèr, and D.~Ippolito, ``Persistent pre-training poisoning of llms,'' 2024. [Online]. Available: \url{https://arxiv.org/abs/2410.13722}
\BIBentrySTDinterwordspacing

\bibitem{shu2023exploitabilityinstructiontuning}
\BIBentryALTinterwordspacing
M.~Shu, J.~Wang, C.~Zhu, J.~Geiping, C.~Xiao, and T.~Goldstein, ``On the exploitability of instruction tuning,'' 2023. [Online]. Available: \url{https://arxiv.org/abs/2306.17194}
\BIBentrySTDinterwordspacing

\bibitem{cole2024memoryinjectionprimer}
M.~Cole, ``Engineer-friendly primer: Memory injection attacks on llms,'' \url{https://murraycole.com/posts/llm-memory-injection-attacks}, 2024, accessed August 2025.

\bibitem{llmsecurity2024memorypoisoning}
L.~S. Database, ``Llm memory poisoning attack,'' Online, 2024, \url{https://www.promptfoo.dev/lm-security-db/vuln/llm-memory-poisoning-attack-01ba0c8d}.

\bibitem{dong2025practicalmemoryinjectionattack}
\BIBentryALTinterwordspacing
S.~Dong, S.~Xu, P.~He, Y.~Li, J.~Tang, T.~Liu, H.~Liu, and Z.~Xiang, ``A practical memory injection attack against llm agents,'' 2025. [Online]. Available: \url{https://arxiv.org/abs/2503.03704}
\BIBentrySTDinterwordspacing

\bibitem{yang2025awesomepoisoning}
P.~Yang, ``Awesome data poisoning and backdoor attacks,'' \url{https://github.com/penghui-yang/awesome-data-poisoning-and-backdoor-attacks}, 2025, accessed: August 2025.

\bibitem{alber2024medicalllm}
M.~Alber, D.~Mack, A.~Ondrus \emph{et~al.}, ``Medical large language models are vulnerable to data-poisoning attacks,'' \emph{Nature Medicine}, vol.~30, no.~6, pp. 1355--1358, 2024.

\bibitem{zhao2025datapoisoningdeeplearning}
\BIBentryALTinterwordspacing
P.~Zhao, W.~Zhu, P.~Jiao, D.~Gao, and O.~Wu, ``Data poisoning in deep learning: A survey,'' 2025. [Online]. Available: \url{https://arxiv.org/abs/2503.22759}
\BIBentrySTDinterwordspacing

\bibitem{fendley2025systematicreviewpoisoningattacks}
\BIBentryALTinterwordspacing
N.~Fendley, E.~W. Staley, J.~Carney, W.~Redman, M.~Chau, and N.~Drenkow, ``A systematic review of poisoning attacks against large language models,'' 2025. [Online]. Available: \url{https://arxiv.org/abs/2506.06518}
\BIBentrySTDinterwordspacing

\bibitem{zou2023universal}
A.~Zou, Z.~Wang, N.~Carlini, M.~Nasr, J.~Z. Kolter, and M.~Fredrikson, ``Universal and transferable adversarial attacks on aligned language models,'' \emph{arXiv preprint arXiv:2307.15043}, 2023.

\bibitem{liu2024autodangeneratingstealthyjailbreak}
\BIBentryALTinterwordspacing
X.~Liu, N.~Xu, M.~Chen, and C.~Xiao, ``Autodan: Generating stealthy jailbreak prompts on aligned large language models,'' 2024. [Online]. Available: \url{https://arxiv.org/abs/2310.04451}
\BIBentrySTDinterwordspacing

\bibitem{jiang2023forcinggenerativemodelsdegenerate}
\BIBentryALTinterwordspacing
S.~Jiang, S.~R. Kadhe, Y.~Zhou, L.~Cai, and N.~Baracaldo, ``Forcing generative models to degenerate ones: The power of data poisoning attacks,'' 2023. [Online]. Available: \url{https://arxiv.org/abs/2312.04748}
\BIBentrySTDinterwordspacing

\bibitem{song2025chainofthoughtpoisoningattacksr1based}
\BIBentryALTinterwordspacing
H.~Song, Y.~an~Liu, R.~Zhang, J.~Guo, and Y.~Fan, ``Chain-of-thought poisoning attacks against r1-based retrieval-augmented generation systems,'' 2025. [Online]. Available: \url{https://arxiv.org/abs/2505.16367}
\BIBentrySTDinterwordspacing

\bibitem{zhao2025shadowcotcognitivehijackingstealthy}
\BIBentryALTinterwordspacing
G.~Zhao, H.~Wu, X.~Zhang, and A.~V. Vasilakos, ``Shadowcot: Cognitive hijacking for stealthy reasoning backdoors in llms,'' 2025. [Online]. Available: \url{https://arxiv.org/abs/2504.05605}
\BIBentrySTDinterwordspacing

\bibitem{su2024enhancingadversarialattackschain}
\BIBentryALTinterwordspacing
J.~Su, ``Enhancing adversarial attacks through chain of thought,'' 2024. [Online]. Available: \url{https://arxiv.org/abs/2410.21791}
\BIBentrySTDinterwordspacing

\bibitem{xiang2024badchainbackdoorchainofthoughtprompting}
\BIBentryALTinterwordspacing
Z.~Xiang, F.~Jiang, Z.~Xiong, B.~Ramasubramanian, R.~Poovendran, and B.~Li, ``Badchain: Backdoor chain-of-thought prompting for large language models,'' 2024. [Online]. Available: \url{https://arxiv.org/abs/2401.12242}
\BIBentrySTDinterwordspacing

\bibitem{snyk_eventstream_2018}
{Snyk Threat Research}, ``Malicious code found in npm package event-stream,'' \url{https://snyk.io/blog/malicious-code-found-in-npm-package-event-stream/}, 2018.

\bibitem{npm_eventstream_2018}
{npm Blog}, ``Details about the event-stream incident,'' \url{https://blog.npmjs.org/post/180565383195/details-about-the-event-stream-incident}, 2018.

\bibitem{shao2025enhancingpromptinjectionattacks}
\BIBentryALTinterwordspacing
Z.~Shao, H.~Liu, J.~Mu, and N.~Z. Gong, ``Enhancing prompt injection attacks to llms via poisoning alignment,'' 2025. [Online]. Available: \url{https://arxiv.org/abs/2410.14827}
\BIBentrySTDinterwordspacing

\bibitem{wang2025promptsafeinvestigatingprompt}
\BIBentryALTinterwordspacing
J.~Wang, P.~Gupta, I.~Habernal, and E.~Hüllermeier, ``Is your prompt safe? investigating prompt injection attacks against open-source llms,'' 2025. [Online]. Available: \url{https://arxiv.org/abs/2505.14368}
\BIBentrySTDinterwordspacing

\bibitem{zhang2024goalguidedgenerativepromptinjection}
\BIBentryALTinterwordspacing
C.~Zhang, M.~Jin, Q.~Yu, C.~Liu, H.~Xue, and X.~Jin, ``Goal-guided generative prompt injection attack on large language models,'' 2024. [Online]. Available: \url{https://arxiv.org/abs/2404.07234}
\BIBentrySTDinterwordspacing

\bibitem{liu2024automaticuniversalpromptinjection}
\BIBentryALTinterwordspacing
X.~Liu, Z.~Yu, Y.~Zhang, N.~Zhang, and C.~Xiao, ``Automatic and universal prompt injection attacks against large language models,'' 2024. [Online]. Available: \url{https://arxiv.org/abs/2403.04957}
\BIBentrySTDinterwordspacing

\bibitem{liu2024promptinjectionattackllmintegrated}
\BIBentryALTinterwordspacing
Y.~Liu, G.~Deng, Y.~Li, K.~Wang, Z.~Wang, X.~Wang, T.~Zhang, Y.~Liu, H.~Wang, Y.~Zheng, and Y.~Liu, ``Prompt injection attack against llm-integrated applications,'' 2024. [Online]. Available: \url{https://arxiv.org/abs/2306.05499}
\BIBentrySTDinterwordspacing

\bibitem{zhang2024studypromptinjectionattack}
\BIBentryALTinterwordspacing
W.~Zhang, X.~Kong, C.~Dewitt, T.~Braunl, and J.~B. Hong, ``A study on prompt injection attack against llm-integrated mobile robotic systems,'' 2024. [Online]. Available: \url{https://arxiv.org/abs/2408.03515}
\BIBentrySTDinterwordspacing

\bibitem{openai_function_calling}
{OpenAI}, ``Function calling and code interpreter,'' \url{https://platform.openai.com/docs/guides/function-calling}, 2023, accessed: 2025-07-28.

\bibitem{schick2023toolformer}
T.~Schick and H.~Schütze, ``Toolformer: Language models can teach themselves to use tools,'' \emph{arXiv preprint arXiv:2302.04761}, 2023.

\bibitem{yao2022react}
S.~Yao, J.~Yang, N.~Yu, D.~Jiang, D.~Wang, and M.~Tang, ``React: Synergizing reasoning and acting in language models,'' \emph{arXiv preprint arXiv:2210.03629}, 2022.

\bibitem{wu2025multiagentautonomousdrivingsystems}
\BIBentryALTinterwordspacing
Y.~Wu, D.~Li, Y.~Chen, R.~Jiang, H.~P. Zou, W.-C. Huang, Y.~Li, L.~Fang, Z.~Wang, and P.~S. Yu, ``Multi-agent autonomous driving systems with large language models: A survey of recent advances,'' 2025. [Online]. Available: \url{https://arxiv.org/abs/2502.16804}
\BIBentrySTDinterwordspacing

\end{thebibliography}

\end{document}